\documentclass[acmsmall]{acmart}
\acmJournal{PACMPL}
\acmVolume{1}
\acmNumber{CONF} 
\acmArticle{1}
\acmYear{2018}
\acmMonth{1}
\acmDOI{} 
\startPage{1}

\setcopyright{none}

\usepackage[utf8]{inputenc}
\usepackage{amsmath}
\usepackage{program}
\usepackage{amsfonts}
\usepackage{listings}
\usepackage{stmaryrd}
\usepackage{wrapfig}
\usepackage{color}
\usepackage{proof}
\usepackage{thmtools,thm-restate}
\usepackage{tabularx}
\title{Veracity: Declarative Multicore\\Programming with Commutativity}
\usepackage{caption}
\usepackage{subcaption}
\usepackage{stackengine}
\usepackage{pifont}
\usepackage[clock]{ifsym}

\usepackage[normalem]{ulem}

\definecolor{codegreen}{rgb}{0,0.6,0}
\definecolor{codegray}{rgb}{0.5,0.5,0.5}
\definecolor{codepurple}{rgb}{0.58,0,0.82}
\definecolor{backcolour}{rgb}{0.97,0.97,0.97}
\lstdefinestyle{vcy}{
    backgroundcolor=\color{backcolour},   
    commentstyle=\color{codegreen},
    keywordstyle=\color{magenta},
    stringstyle=\color{codepurple},
    basicstyle=\ttfamily\scriptsize,
    breakatwhitespace=false,
        numbers=none,
        numbersep=1pt,
        frame=lines,
        rulecolor=\color{codepurple},
    breaklines=true,                 
    captionpos=b,                    
    keepspaces=true,                                
    showspaces=false,                
    showstringspaces=false,
    showtabs=false,                  
    tabsize=2,
        mathescape=true,
        escapechar=|
}
\lstdefinelanguage{VCY}[]{C}{
    morekeywords={commute}
}

\newenvironment{itemize*}%
  {\begin{itemize}%
    \setlength{\itemsep}{0.0in}%
    \setlength{\topsep}{0.0in}%
    \setlength{\parskip}{0.0in}}%
  {\end{itemize}}
\newenvironment{enumerate*}%
  {\begin{enumerate}%
    \setlength{\itemsep}{0.0in}%
    \setlength{\topsep}{0.0in}%
    \setlength{\parskip}{0.0in}}%
  {\end{enumerate}}

\newcommand\sem[1]{\llbracket #1 \rrbracket}
\newcommand\ttt[1]{\texttt{#1}}
\newcommand\ttx{\texttt{x}}
\newcommand\tty{\texttt{y}}
\newcommand\ttz{\texttt{z}}

\newcommand\ttc{\texttt{c}}
\newcommand\ttcommute{\texttt{commute}}

\newcommand\block[1]{\{#1\}}
\newcommand\angles[1]{\langle#1\rangle}
\newcommand\sskip{\text{skip}}

\newcommand\execution{\varepsilon}
\newcommand\transition{\ell}
\newcommand\eg{{\it e.g.}}
\newcommand\ie{{\it i.e.}}

\newcommand\frag{\ensuremath{\mathfrak{f}}}
\newcommand\fragA{\ensuremath{\frag_1}}
\newcommand\fragB{\ensuremath{\frag_2}}
\newcommand\fragC{\ensuremath{\frag_3}}

\newcommand\benchCount{30}

\newtheorem{prop}{Proposition}

\author{Adam Chen}
\affiliation{\institution{Stevens Institute of Technology} \country{USA}}
\email{achen19@stevens.edu}
\author{Parisa Fathololumi}
\affiliation{\institution{Stevens Institute of Technology} \country{USA}}
\email{pfatholo1@stevens.edu}
\author{Eric Koskinen}
\affiliation{\institution{Stevens Institute of Technology} \country{USA}}
\email{eric.koskinen@stevens.edu}
\author{Jared Pincus}
\affiliation{\institution{Stevens Institute of Technology} \country{USA}}
\email{jpincus@stevens.edu}

\begin{document}
\begin{abstract}
There is an ongoing effort to provide programming abstractions that ease the burden of exploiting multicore hardware. Many programming abstractions 
(\eg, concurrent objects, transactional memory, etc.) simplify matters, but still involve intricate engineering.
We argue that some difficulty of multicore programming can be meliorated through a declarative programming style in which programmers directly express the independence of fragments of sequential programs.

In our proposed paradigm, programmers write programs in a familiar, sequential manner, with the added ability to explicitly express the conditions under which code fragments sequentially commute. 
Putting such commutativity conditions into source code 
offers a new entry point for a compiler to
exploit the known connection between commutativity and parallelism.
%
We give a semantics for the programmer's sequential perspective
and, under a correctness condition, find that a compiler-transformed parallel execution is equivalent to the sequential semantics.
Serializability/linearizability are not the right fit for this condition, so 
we introduce
scoped serializability and show how it can be enforced with lock synthesis techniques.

We next describe a technique for automatically verifying and synthesizing commute conditions  via a new reduction from our commute blocks to logical specifications, upon which symbolic commutativity reasoning can be performed.
We implemented our work in a new language called Veracity, implemented in Multicore OCaml.
%
We show that commutativity conditions can be automatically generated 
across a variety of new benchmark programs,
confirm the expectation that concurrency speedups can be seen as the computation increases, and apply our work to a small in-memory filesystem and an adaptation of a crowdfund blockchain smart contract.

\end{abstract}

\title{Veracity: Declarative Multicore Programming with Commutativity}         

\maketitle


\section{Introduction}

%

Writing concurrent programs is difficult. Researchers and practitioners, seeking to make life easier, have developed better paradigms for concurrent programming such as concurrent objects~\cite{hw}, 
transactional memory~\cite{HerlihyM1993,dstm,HarrisF2003,intelstm},
 actors~\cite{Armstrong1997}, parallel for~\cite{fortress}, goroutines~\cite{Prabhakar2011}, ownership~\cite{rust}, composable atomicity~\cite{Golan2015}, etc.
As another strategy, compiler designers have instead sought to automatically parallelize programmers' sequential programs
(see, \eg~\cite{Li1990,Blume1992,Blume1994}). 
%
Others have devised more \emph{declarative} and/or domain-specific programming models
such as Fortress~\cite{Steele2005}. Some have extended concurrent collection programming with graph-based languages for GPUs~\cite{Grossman2010};
some have introduced DSLs for grid computing~\cite{Orchard2010};
some delay side effects in blocks~\cite{Lindley2007};
others aim to further separate high-level programming from low-level concurrent programming (\eg~the Habanero project~\cite{Barik2009}).
(See also the DAMP Workshop\footnote{Proceedings of the Workshop on Declarative Aspects of Multicore Programming (DAMP), 2007--2012.}.) 
Still others aim to improve proof techniques (\eg~\cite{Owicki1976,Brookes2016,Elmas2010,Vafeiadis2010}) so that verification tools can complement the error-prone task of concurrent programming.

Meanwhile, it has been long known that \emph{commutativity} can be used for scalability in concurrent programming, dating back to the database community~\cite{weihl,weihlcommu,commu,korth} and, more recently, for system building~\cite{Clements2015}. There have thus been a growing collection of works aimed at exploiting commutativity~\cite{Kulkarni2007,Kulkarni2008,ont,ppopp08,Kulkarni2011,Hassan2014,Dickerson2017,Dickerson2019},
synthesizing/verifying commutativity~\cite{aleen,KR:PLDI11,DBLP:conf/cav/GehrDV15,Bansal2018,Koskinen2021},
and using commutativity analysis for parallelization in compilers~\cite{rinard} and blockchain smart contracts~\cite{Pirlea2021}.

Despite works on declarative programming models for concurrency and separate works that exploit commutativity, to our knowledge none have yet sought to combine the two by giving the programmer (or automated analysis) agency in specifying commutativity.

\paragraph{Programming with commutativity.}
In this paper, we introduce a new sequential programming paradigm in which \emph{explicit} commutativity conditions are part of the language, to be exploited by the compiler for concurrent execution. Specifically, we introduce \ttcommute{} statements of the form:
\[
\texttt{commute (expr) \block{\{ stmt$_1$ \} \{ stmt$_2$ \}}}
\]
The idea is to allow programmers to continue to write programs using sequential reasoning and, still from that sequential perspective,  express the conditions 
under which statements commute.
For example, a programmer may specify that statements
incrementing and decrementing a counter \ttc{} commute when the counter's value is above 0. 
Putting commutativity directly into the language has three key benefits:
\begin{enumerate*}
\item \emph{Sequential programming}: Programmers need only reason sequentially, which is far more approachable than concurrent reasoning. The semantics are simply that
non-deterministic behavior is permitted when \ttcommute{} conditions hold and, otherwise, resort to sequencing in the order written.
\item \emph{Sequential verification/inference}: These commutativity conditions can be automatically verified or even synthesized without the need to consider all interleavings.
\item \emph{Parallel execution}: Once commutativity conditions are available, they can be used \emph{within the compiler} and combined with lock synthesis techniques (\eg~\cite{gulwani,Vechev2010} and some new ones) to generate parallel compiled programs.
\end{enumerate*}


Since programmers write with the sequential semantics in mind, we naturally would like a concurrent semantics that is equivalent to the sequential semantics but, interestingly, the {\it de facto} standards of 
serializability~\cite{Papadimitriou1979} and 
linearizability~\cite{hw} do not quite fit the bill. Parallelizing programs with arbitrarily nested (and sequentially composed) \ttcommute{} blocks, involves a recursive sub-threading shape.
We introduce \emph{scoped serializability} to describe a sufficient condition under which the concurrent semantics is equivalent to the sequential semantics, when used with valid commutativity conditions. Scoped serializability is \emph{stronger} than serializability; though that may at first seem distasteful, it is necessary to capture the structure of nested commute blocks, and it is not bad given that the program can be written by only thinking of its sequential semantics in contrast to, say, transactional memory where the programmer must decide how to organize their program into threads and atomic sections.
Scoped serializability can then be enforced through lock synthesis techniques, another area of ongoing research (\eg~\cite{Flanagan2003,gulwani,Vechev2010,Golan2015}). While such existing techniques can be applied here, they are geared to a more general setting. We describe alternative locking and re-write strategies that incur less mutual exclusion and are more geared toward scoped serializability.

Although \ttcommute{} conditions can be written by programmers, we next focus on correctness and describe how such conditions can be automatically verified and even inferred. We describe a novel translation from \ttcommute{} programs to an embedding as a logical specification, for which recent techniques~\cite{Bansal2018} can verify or synthesize commutativity conditions. The pieces don't immediately align: \ttcommute{} blocks occur in program contexts (with local/global variables) and our language provides support for nested commute blocks as well as builtin primitives such as dictionaries. We show that, nonetheless, these program pieces can be translated to logical specifications, which can be used 
to synthesize \ttcommute{} conditions via abstraction-refinement~\cite{Bansal2018}.
Synthesized conditions can sometimes be overly complex and/or describe only trivial cases (due to incompleteness in those algorithms) so
a programmer may instead wish to manually-provide a \ttcommute{} condition which we then verify.

We implement our work in a front-end compiler/interpreter for a new language called Veracity\footnote{A portmanteau of ``verified'' and ``commutativity.''}, built on top of Multicore OCaml~\cite{multicoreOcaml} 4.12.0, and a new commutativity condition verifier/synthesizer, using a variety of underlying SMT solvers.
Veracity uses the recent highly-concurrent hashtable implementation \ttt{libcuckoo}~\cite{libcuckoo}.
We will publicly release Veracity, as well as \benchCount{} benchmark programs which are, as far as we know,
the first to include \ttcommute{} blocks.
We provide a preliminary experimental evaluation, offering promising evidence that (i) \ttcommute{} conditions can be automatically verified/inferred, (ii) as expected, speedups can be seen as the computation size grows, and (iii) \ttcommute{} blocks could be used in applications such as in-memory file systems and blockchain smart contracts.


\paragraph{Contributions.} In summary, our contributions are:
\begin{itemize}
\item \ttcommute{} statements, a new sequential language primitive that weakens sequential composition, to explicitly express commutativity conditions, which are then used for parallelization.

\item (Sec.~\ref{sec:semantics}) Sequential and concurrent semantics for the language.

\item (Sec.~\ref{sec:scoped-serializability}) A correctness condition for parallelization called \emph{scoped serializability} that, with commutativity, implies that the concurrent semantics are equal to the sequential semantics. We also present methods for automatically enforcing scoped serializability.

\item (Sec.~\ref{sec:verifyinfer}) A method for automatic verification and synthesis of \ttcommute{} conditions, via an embedding into ADT specifications.

\item (Sec.~\ref{sec:impl}) A frontend compiler/interpreter for a new language Veracity, 
built in Multicore OCaml with
libcuckoo
and a new implementation of Servois~\cite{servois}.

\item (Sec.~\ref{sec:eval}) A preliminary evaluation demonstrating synthesized \ttcommute{} conditions, scaling speedups, and relevant applications.

\end{itemize}
\noindent
\emph{Our implementation of Veracity and all benchmark programs can be found in the supplemental materials. Benchmark sources are also given in Apx.~\ref{apx:benchmarks}.}

\paragraph{Limitations.} While we have implemented a multicore interpreter, we reserve back-end compilation questions to future research: there is much to be explored there, particularly with respect to optimization. 
When commute blocks have loops in them, we infer/verify commute conditions by first instrumenting loop summaries~\cite{Ernst2020,Kroening2008,Xie2017,Silverman2019} with havoc/assume. We used Korn\footnote{\url{https://github.com/gernst/korn}} and some manual reasoning to increase Korn's precision, but we plan to fully automate loop support in the future.
%
%
Scoped serializability is formalized in Sec.~\ref{subsec:enforcing} and we argue that it can be automatically enforced.
Because automated lock synthesis is largely an orthogonal research area, not a core contribution of our work, and would involve implementing other orthogonal components (such as alias analysis), our prototype implementation does not perform lock synthesis automatically. Rather, we manually applied the procedure of Sec.~\ref{subsec:enforcing} and detail how it applied to our benchmarks in Sec.~\ref{sec:eval}.
 

%

\section{Overview}
\label{sec:overview}

We begin with some simple examples
to illustrate how a programmer may use \texttt{commute} statements and, from these examples, outline the challenges addressed in this paper. The following is perhaps the most trivial example: commuting blocks that increment and decrement a value.
\begin{center}
\begin{program}[style=tt]
commute(true) \{ \{ c = c-x; \}  \{ c = c+y; \} \}
\end{program}
\end{center}
\noindent
These two operations clearly commute: the final value of \ttc\ is the same regardless of the order in which the blocks are executed, due to the natural commutativity of integer arithmetic. Moreover, this holds for any initial value of \ttc, so we have used the trivial \emph{commutativity condition} \texttt{true}. Compiler optimizations can exploit simple situations like this where commutativity always holds and parallelize with a lock to protect the data race~\cite{rinard}. In this example, the overhead of parallelization is not worth it, so let us move to more examples.
Now consider a case where the condition is not simply \texttt{true}, and must be specified.

\begin{example}[Conditional commutativity - \tt{simple.vcy}]\label{ex:simple} 
\begin{lstlisting}[style=vcy,language=VCY]
commute (c>a) { 
  |\fragA|: { t = foo(c>b); a = a - $\mid$t$\mid$; } 
  |\fragB|: { u = bar(c>a); }
}
\end{lstlisting}
\end{example}
\noindent

Above
we refer to the two blocks labeled \fragA,\fragB\ as the two ``fragments'' (or co-fragments) of the \ttcommute{} statement\footnote{The terminology is chosen to avoid nebulousness words like ``blocks,'' ``statements,'' and ''nested.''}.
We assume that \texttt{foo} and \texttt{bar} are pure but costly computations.
In this case, the programmer has written the commute condition \texttt{c>a}. Although \fragB\ reads the variable \texttt{a} which is written by \fragA, it only observes whether or not \texttt{c>a}. Moreover, \fragA\ may modify \texttt{a}, but will only decrease its value, which does not impact the boolean observation \ttt{c>a} made by \fragB. Hence, the fragments commute whenever \texttt{c>a} initially. 
It is necessary to include a condition
because when the condition does not hold, the operations may not always commute
(they only commute if \ttt{foo} returns a value with magnitude greater than \ttt{a-c}, which may be computationally difficult to determine \emph{a priori})
and we must resort to sequential execution.
Note that, throughout this paper, we mostly use \ttcommute{} statements with only two fragments although our implementation supports $N$-ary \ttcommute{} statements.

We emphasize that a programmer needs to reason only in the sequential setting. However, as detailed below, our compiler can transform the above program into a concurrent one, permitting \texttt{foo} and \texttt{bar} to execute in parallel when \texttt{c>a} and resort to sequential execution otherwise. Note that it is not possible 
to parallelize these computations with a simple dataflow analysis.  The challenge is that \texttt{bar(c>a)} cannot be moved above \texttt{a = a - $\mid$t$\mid$} in the absence of protection from the \ttcommute{} condition. 
More generally, the intuition is that each fragment may make observations about the state and also perform mutations to the state. These fragments can be parallelized if the mutations made by one do not change the observations made by the other, which is captured by the notion of commutativity and specified in the \ttt{commute} expression.
Furthermore, the compiler does not need to introduce any locks because there is a read-write in one fragment, but only a single read in the other, and the \ttt{commute} condition ensures that reading before or after the write does not matter.


\begin{example}[Counter control flow - \texttt{calc.vcy}]\label{ex:calc}
\begin{lstlisting}[style=vcy,language=VCY]
commute (c>0) { 
  |\fragA|: { x = calc1(a); c = c + (x*x); }
  |\fragB|: { if (c>0 && y<0) 
         { c = c-1; z = calc2(y); }
       else
         { z = calc3(y); } }
}
\end{lstlisting}
\end{example}
\noindent
This example considers \ttcommute{} blocks with fragments that have control flow. Below
\ttt{calc1}/\ttt{calc2}/\ttt{calc3}  are again pure calculations.
Here, again using only sequential reasoning, the programmer has provided the \ttt{commute} condition \texttt{c>0}. This is necessary because, \eg, if \ttc\ is 0 initially, then in one order \ttc\ will be \texttt{x*x} and in the other order, \ttc\ will be \texttt{(x*x)-1}. 
%
A compiler can use this sequential commutativity condition to parallelize these operations, provided that it also ensures a new form of serializability (detailed in the next subsection) through lock synthesis techniques.
In this case, a single lock must be synthesized
to protect access to \ttc{}, yet allow parallel execution of \texttt{calc1} with \texttt{calc2}/\texttt{calc3}
(\ie, \fragA\ holds the lock only during the \ttc{} read/write, and \fragB\ holds the lock from the beginning until before the call to \texttt{calc2}/\texttt{calc3}).
This enforces serializability of the fragments, so we may treat them as atomic points in the execution, and then, due to the \ttcommute{} condition, it does not matter which point occurs first.

This example also shows that other orthogonal strategies such as future/promises~\cite{Liskov1988,Chatterjee1989} do not subsume \ttcommute{} blocks. A promise for the value of \ttx\ must be evaluated at the end of \fragA\ before beginning \fragB\ because \ttx\ is used in \fragA. 
Promises cannot avoid the data-flow dependency across \fragA\ into \fragB. By contrast commutativity allows a compiler to execute the fragments out of order, effectively relaxing the data dependency
(\ie~although \ttc{} may change, \texttt{c>0} will not).
(In future work one could combine promises with \ttcommute{} blocks.)  

Commutative computation appears in many contexts that go beyond counters, \eg, vector or matrix multiplication: 
\begin{example}[Multiplication - \tt{matrix.vcy}]\label{ex:matrix}
\begin{lstlisting}[style=vcy,language=VCY]
commute (y == 0) {
   { int sc = scale(y); int y|$_0$| = y;
     x = x*sc; y = 3*y; z = z - 2*y|$_0$|; }
   { int y|$_0$| = y;
     x = 5*x; y = 4*y; z = 3*z - y|$_0$|; 
     out = summarize(z); }}
\end{lstlisting}

\end{example}
\noindent
Here the resulting vector (\ttx,\tty,\ttz) can be the same in either order of the blocks, but only in certain cases.
This is almost a trivial diagonal matrix multiplication, except that there is a non-diagonal contribution to \ttz. Under the condition that \tty=0 initially, however, the contribution vanishes.
Our compiler can exploit this commutativity and execute these fragments in parallel, synthesizing a lock for the read/writes to \ttx{}/\tty/\ttz{}.
If the \ttt{scale} and \ttt{summarize} computations are time-consuming then there is a large payoff for parallelizing: locks are held for only small windows, and execution of the costly \texttt{scale} and \texttt{summarize} can overlap.

It is increasingly common for programming languages to have \emph{builtin abstract datatypes} (ADTs), such as dictionaries/hashtables, queues, stacks, etc. These present further opportunities for sequential commutativity-based programming and automatic parallelization through highly concurrent implementations of those ADTs. Consider the following example:
\begin{example}[Builtin dictionaries - \tt{dict.vcy}]\label{ex:dict} 
\begin{lstlisting}[style=vcy,language=VCY]
commute (res $\neq$ input) {
  |\fragA|:  { t = calc(x); stats[res] = t; } 
  |\fragB|:  { y = stats[input]; y = y + x; }
}
\end{lstlisting}
\end{example}
\noindent
Above, \ttt{stats} is a dictionary/hashtable, with a \ttt{put} operation in fragment \fragA, and a \ttt{get} in \fragB. The programmer (or our analysis) has provided a \ttcommute{} condition that the keys are distinct, using sequential reasoning.

Hashtables have appealing commutativity properties~\cite{ont,ppopp08,Bronson2010}, well understood sequential semantics and recent high-performance concurrent implementations~\cite{Li2014,Liu2014} such as \texttt{libcuckoo}~\cite{libcuckoo}.
For this example, the memory accesses do not conflict  except on the hashtable. Thus, the compiler can translate the \ttcommute{} block to be 
executed concurrently simply by using a hashtable with a concurrent implementation like  \texttt{libcuckoo}, with no other locking needed.
That linearizable implementation ensures the atomicity of the operations on \ttt{stat} and the \ttcommute{} condition (\ttt{res $\neq$ input}) ensure that \fragA\ will commute with \fragB\ regardless of the interleaving.
\ttcommute{} blocks that use other builtin ADTs such as Sets, Stacks or Queues can similarly benefit from this strategy
(locking may be necessary when there are multiple such operations in each fragment).

As seen in the examples above, our work is particularly suited to settings where there are mixtures of long pure computation, mixed with some commutative updates to shared memory. Examples might include in-memory filesystems, machine learning, and smart contracts.
In this paper we focus on the theoretical foundations, algorithms, and verification/synthesis with some common-sense empirical results, but defer real-world applications and integration to future work.

%

\subsection{Semantics of \ttcommute}
\label{subsec:overview-semantics}

Our proposed language has no explicit \texttt{fork} or \texttt{clone} command and so a user cannot explicitly express concurrency. Instead, a user (and, later, our automated techniques) can express \emph{sequential} nondeterminism of code fragments through \ttcommute{} statements. 
Parallelism is only introduced by our compiler, which exploits the \ttcommute{}-specified allowance of nondeterminism.
Incorporating such \ttcommute{} statements into a programming language has semantic implications.
A natural goal is to provide a concurrent semantics that somehow corresponds to a sequential semantics, \eg\ trace inclusion, simulation, etc.
%
However, the programming model is also distinct from many prior models that involve explicit fork/clone, like transactional memory or concurrent objects and their
respective correctness conditions:
serializability/opacity~\cite{Papadimitriou1979,opacity}
or linearizability~\cite{hw}. 

Because of this, there is also a distinction in what correctness condition is appropriate.
To see the distinction from transactional memory and serializability, consider the following \ttcommute\ blocks:
\begin{example}[Nested commute blocks -- \tt{nested.vcy}]\label{ex:notser} 
\begin{lstlisting}[style=vcy,language=VCY]
y := 0; x := 1;
commute(true) {
   { commute(true) 
      |\fragA|: { x := 0; }
      |\fragB|: { x := x * 2; } } 
 |\fragC|:{ if (x>2) y := 1; }
}
\end{lstlisting}
\end{example}
\noindent
A notion of serializability adapted for this setting might require that any execution be equivalent to some serial order of  $\fragA{}$, $\fragB{}$, and $\fragC{}$. This would permit the interleaved execution:
\[
\!\texttt{y:=0},\;\; 
\texttt{x:=1},\;\; 
\fragB\!:\!\texttt{x:=x*2},\;\; 
\fragC\!:\!\texttt{y:=1},\;\; 
\fragA\!:\!\texttt{x:=0}\]
ending in a final state where \ttx=0 and \tty=1. However, the semantics of \ttcommute{} in our language does not permit a serial order where $\fragC{}$ occurs between ($\fragA{}$;$\fragB{}$) nor between ($\fragB{}$;$\fragA{}$).

In Section~\ref{sec:semantics} we introduce sequential and concurrent semantics for a simple language augmented with \ttcommute{} statements. From the programmer's perspective, the semantics of \ttcommute{} statements are naturally captured through nondeterminism.
At runtime however, our interpreter uses a parallel execution semantics through recursively nested threading.
We define a correctness condition called
\emph{scoped serializability} which is stronger than serializability yet ensures an equivalence between the concurrent semantics and the sequential semantics.
This equivalence means that the program can be automatically parallelized.
We then give some techniques for \emph{enforcing} scoped serializability.
One \emph{can} na\"{i}vely use mutual exclusion or existing lock synthesis techniques (which were designed for programs with explicit forking). We then discuss improved locking techniques that are more geared to this nested, pair-wise setting.
We show that, in some cases, some sound reordering of fragments can minimize or even eliminate the need for locking.


\subsection{Automated Verification and Inference of \ttcommute{} Conditions}

A key benefit of \ttcommute{} statements is that we can use \emph{sequential reasoning} to discover commutativity conditions (\ie~we do not have to consider interleavings). 
Commute conditions are often simple enough for programmers to write, though they must be correct in order for parallel execution to be safe. In Sec.~\ref{sec:verifyinfer} we turn to this correctness issue and discuss how to automatically verify or even infer these conditions.

A collection of recent techniques and tools have emerged pertaining to commutativity reasoning.
\citet{Bansal2018} used a counterexample-guided abstraction refinement (CEGAR) strategy in the Servois tool~\cite{servois} to synthesize commutativity conditions of object methods' pre/post specifications, but not the {\it ad hoc} code that appears in our \ttcommute{} blocks.
\citet{Koskinen2021} verified commutativity conditions of source code through reachability. We found that in this setting, the reachability approach introduced many unnecessary, extraneous paths and variables (reducing performance and complicating conditions), and that SMT solvers struggled in many cases.
Finally, \citet{Pirlea2021} described an abstract interpretation that tracks the linearity of assignments to determine when operations \emph{always} commute. Yet our \ttcommute{} blocks need \emph{conditions} for commutativity and have also non-linear behaviors such as matrix multiplication, modulus, calls to builtin ADTs, etc.


We enable automated verification or inference of \ttt{commute} conditions through a novel embedding of \ttcommute{} block fragments into logical specifications, fit for verification with SMT or inference with the abstraction-refinement of \citet{Bansal2018}.
The translation is not immediate: \ttcommute{} statements occur within the context of a program, mutate global/local variables, invoke builtin ADTs such as dictionaries, and may involve nested \ttcommute{} statements. 
The idea is thus to capture the program context for the given \ttt{commute} block as the pre-state, and then translate each fragment's code into a logical post-condition, describing how the fragment mutates the context. Builtin operations are treated by ``inlining'' their specification and nested \ttcommute{} blocks can be treated as sequential composition, thanks to their sequential semantics.
Synthesized \ttcommute{} conditions can then be reverse translated back into the source language.
Sometimes these conditions may be overly complex or too trivial and thus a programmer still may wish to provide their own, which may be verified via the same embedding.

%

\subsection{The Veracity Programming Language}

We implemented our work in a soon-to-be-released prototype front-end compiler and interpreter for a new language called Veracity, built on top of Multicore OCaml~\cite{multicoreOcaml} and a new implementation of the \citet{Bansal2018} algorithm. Our interpreter performs the embedding to verify and synthesize \ttcommute{} conditions and uses Multicore OCaml domains and a foreign function interface to \texttt{libcuckoo}~\cite{libcuckoo} for concurrent execution of \ttcommute{} statements.

We create a suite of \benchCount{} benchmark programs with \ttcommute{} blocks, which we will also  release publicly. We first show that Veracity can synthesize commutativity conditions for most of these programs, which include a variety of program features: linear and non-linear, vector multiplication, builtin ADTs, modulus, nesting, arrays and loops.
When these synthesized conditions were too complex, we showed that better manual ones could be verified.

We next confirmed the expectation that concurrent execution offers a speedup over sequential execution, and that speedup increases as non-conflicting computations increases. Although no existing programs are written with \ttcommute{} statements, we examined two case studies in which we adapted programs to use \ttcommute{} statements, including an Algorand smart contract and an in-memory file system.

\section{Preliminaries}


We provide some basic background and definitions used throughout this paper.

\emph{States.}
A state $\sigma \in \Sigma$ is a partial mapping $Varnames \sqcup MemLocs \rightharpoonup V$, \ie~the disjoint union of variable names and memory locations, to program values. We assume a notion of scope, and that each program variable statically refers to a unique element of the state. We assume that states can be composed and decomposed by scope, \ie, $\sigma = \sigma_0 \oplus \sigma_1$, and that relations on $\Sigma$ are also relations over the decomposed states. We work with equality on states $\simeq$ in the usual way (\eg, equivalence of valuation of accessible, primitive variables).

%
%


\emph{Syntax.} 
We begin with syntax and contextual semantics for a standard programming language. The language involves constants $c$, references, mutable variables and declarations:
\[\begin{array}{llllll}
	c &::=& int \mid string \mid bool \mid () \mid ref  \hspace{0.5in} \; & v\text{-}decls &::=& \epsilon \mid v\text{-}decls' \\
	lval &::=& \text{varname} \mid ref & v\text{-}decls' &::=& t \text{ } v = e \mid t \text{ } v = e, v\text{-}decls'
\end{array}\]

\noindent
We next define redexes in the usual way:
	\[\begin{array}{lll}
	r &::=& 
	v \mid \text{deref } p \mid c_0[c_1] \mid \texttt{new }t[c] 
	 \mid \texttt{new hashtable}[t_0,t_1] \mid \text{uop } c \mid c_0 \text{ bop } c_1 \\
	&&\mid \texttt{true} ? e_0 : e_1 \mid \texttt{false} ? e_0 : e_1 \mid c.\text{fieldname} 
	\mid lval = c \mid t\text{ }v = c \mid \texttt{while}(e) \block{s} \mid \sskip; s  \\
	&& \mid \texttt{if(true)}\block{s_0}\texttt{else}\block{s_1} 
	\mid \texttt{if(false)}\block{s_0}\texttt{else}\block{s_1} 
	 \boxed{\mid \texttt{commute(false)}\block{\block{s_0} \block{s_1}}}
	\end{array}\]
Our language includes typical constructors, arrays, structures, as well as ``builtin'' ADTs.
(Our theory applies to general ADTs, but for illustration here we use hashtables.)
The new feature of the language is the \ttcommute{} statement, which we discuss in the next section.
All semantics have a \ttt{commute(false)} redex, which is evaluated sequentially. \ttt{commute(true)} is semantics-specific; see Sec.~\ref{sec:semantics}.
We next define contexts in the usual way.
The notation $H[r]$ means ``$H$ with the $\bullet$ replaced by $r$''.
	\[\begin{array}{lll}
		H &::=& \bullet \mid H ; s \mid \text{uop }H \mid c \text{ bop } H \mid H \text{ bop } e \mid H ? e_0 : e_1\\
		&& \mid H[e] \mid c[H] \mid 
		H.\texttt{fieldname} \mid lval = H \mid t \text{ } v = H\\
		&& \mid \texttt{if}(H) \block{s_0} \texttt{else} \block {s_1}  \mid  \boxed{\texttt{commute}(H) \block{\block{s_0} \block{s_1}}}
	\end{array}\]


\emph{Small step semantics.}
Redex reduction rules are denoted
$\angles{r, \sigma} \leadsto \angles{r', \sigma'}$ and, for lack of space, given in Apx~\ref{apx:redexreductions}. 
For simplicity, these are all atomically reducible expressions. Note that memory cannot be read and written in one atomic step.
For each reduction, we  use the Redex rule below:
\[\infer[\text{Redex}]{\angles{r, \sigma} \leadsto \angles{r', \sigma'}}
\;\;\;\;
\infer[\text{SS}]{\angles{H[r], \sigma} \leadsto \angles{H[r'], \sigma'}}
{\angles{r, \sigma} \leadsto \angles{r', \sigma'}}
{}\]
For a program $s = H[r]$, a small step reduction is an application of the SS rule above.
If $s$ is a redex, we recover the redex rule by taking $H = \bullet$.

Suppose we are given program statements $s_0$ and $s_1$, and a predicate on states $\varphi^{s_0}_{s_1} : \Sigma \rightarrow \mathbb{B}$. 

\begin{definition}[Sufficient Commutativity Condition]
	\label{Sufficient Commutativity Condition}
	$\varphi^{s_0}_{s_1}$ is a sufficient commutativity condition for $s_0$ and $s_1$ when
\[
\begin{array}{ll}
		\forall \sigma_0, \sigma_{f_0}, \sigma_{f_1}: & \angles{s_0; s_1, \sigma_0} \leadsto^* \angles{skip, \sigma_{f_0}} \wedge \\ 
		& \angles{s_1; s_0, \sigma_0} \leadsto^* \angles{skip, \sigma_{f_1}} \implies \\
		\multicolumn{2}{r}{\;\;\;\;\varphi^{s_0}_{s_1}(\sigma_0) = \texttt{true} \implies \sigma_{f_0} = \sigma_{f_1}}
\end{array}
\]
\end{definition}
\noindent
That is $\varphi^{s_0}_{s_1}$ tells when the order of execution has no effect on the final state. 

\section{Semantics of Commute Blocks}
\label{sec:semantics}

In this section we define semantics for the language, which has been extended with the addition of \ttcommute{} statements.
We define three semantics, denoted
$sem ::= seq \mid nd \mid par$.
We explain the main rules below, with some details and reductions deferred to Apx.~\ref{apx:omitted}.


\paragraph{Sequential semantics.} For  sequential semantics $\leadsto_{seq}$ in the language extended with \ttcommute{} blocks,  we add a simple reduction that treats \ttcommute{} as sequential composition:
\[\begin{array}{lll}
	\angles{\texttt{commute(true)}\block{\block{s_0},\block{s_1}}, \sigma} & \leadsto_{seq} & \angles{s_0; s_1, \sigma}
\end{array}\]


\paragraph{Nondeterministic semantics.} 
The nondeterministic semantics $\leadsto_{nd}$ has two possible reductions that could take place when a \ttcommute{} condition has been reduced to \texttt{true}:
\[\begin{array}{lll}
	\angles{\texttt{commute(true)}\block{\block{s_0},\block{s_1}}, \sigma} & \leadsto_{nd} & \angles{s_0; s_1, \sigma} \\
	\angles{\texttt{commute(true)}\block{\block{s_0},\block{s_1}}, \sigma} & \leadsto_{nd} & \angles{s_1; s_0, \sigma} \\
\end{array}\]
The two rules for \texttt{commute(true)} apply to the same syntax. This allows for non-determinism when reducing this redex. We also introduce an SS-Nd rule akin to SS (see Apx.~\ref{apx:omitted}).

\paragraph{Concurrent semantics.} The next semantics permits the bodies of (possibly nested) \ttcommute{} blocks to execute concurrently. To this end, we define a \emph{configuration} $\mathfrak{C}$ that expresses the nested threading nature of the semantics. Formally, there are two constructors of a configuration:
\[
\begin{array}{lll}
	\mathfrak{C} &::=& \angles{s, \sigma} \;\; \mid \;\; \angles{(\mathfrak{C}_0, \mathfrak{C}_1), s, \sigma} 
\end{array}
\]
Configurations are either a top-most configuration (code and state), or a nested configuration, executing in the context of an outer remaining code $s$ and outer state $\sigma$.
The statements of a \ttcommute{}, which we will call \emph{fragments}, execute code that naturally has access to variables defined in outer scopes. Thus when fragments step, they need to access to outer scopes. To allow for this, we define appending a state to a configuration ($\mathfrak{C} \oplus \sigma$) as:
$$
\angles{s, \sigma_0} \oplus \sigma ::= \angles{s, \sigma_0 \oplus \sigma}
$$
$$
\angles{(\mathfrak{C}_0, \mathfrak{C}_1), s, \sigma_0} \oplus \sigma ::= \angles{(\mathfrak{C}_0, \mathfrak{C}_1), s, \sigma_0 \oplus \sigma}
$$
For convenience, as also define $\mathfrak{C}.st$, the state of a configuration, by:
$	\angles{s, \sigma}.st = \sigma $
and
$	\angles{(\mathfrak{C}_0, \mathfrak{C}_1), s, \sigma}.st = \sigma$.

Nested configurations arise when reaching a \ttcommute{} block. That is, start with the following redex and corresponding reduction:
\[
\begin{array}{c}
	\angles{\texttt{commute(true)} \block{\block{s_0}, \block{s_1}}
	} \leadsto_{par} \angles{(\angles{s_0, \emptyset}, \angles{s_1, \emptyset}), \sskip, \sigma} 
	\end{array}
\]
where $\emptyset$ represents new local (and empty) scopes of variables for each fragment. 
This is then included in the $par$ semantics with the following rule:
$$
\infer[\text{Fork-Step}]
{\angles{H[r], \sigma} \leadsto_{par} \angles{(\angles{s_0, \sigma_0}, \angles{s_1, \sigma_1}), H[r'], \sigma}}
{\angles{r, \sigma} \leadsto_{par} \angles{(\angles{s_0, \sigma_0}, \angles{s_1, \sigma_1}), r', \sigma}}
$$

Execution in the concurrent semantics involves (possibly recursively) descending into sub-components to find a leaf fragment that can take a step. This is done using the following Left- (or, similarly, Right-) projection rule:
\[\infer[\stackanchor{Left-Proj}{(mut. mut.) R-Proj}]
{\angles{(\mathfrak{C}_0, \mathfrak{C}_1), s, \sigma} \leadsto_{par} \angles{(\mathfrak{C}_0', \mathfrak{C}_1), s, \sigma'}}
{\mathfrak{C}_0 \oplus \sigma \leadsto_{par} \mathfrak{C}_0' \oplus \sigma'}
\]

When the two fragments of a \ttcommute{} have both reduced their statements to \sskip, the Join rule can be used. Updates to the inner scope are lost, but updates to the outer scope $\sigma_2$ are preserved:
\[\infer[\text{Join}]
{\angles{(\angles{\sskip, \sigma_0}, \angles{\sskip, \sigma_1}), s, \sigma_2} \leadsto_{par} \angles{s, \sigma_2}}
{}\]
\noindent
Note that join is blocking; both threads must reach \sskip{} before $s$ begins to reduce. 

\paragraph{Executions.}
An \emph{execution} $\execution$ (w.r.t. a semantics $sem$) is a sequence of configurations with labeled transitions. That is, let $\mathfrak{C}$ be the set of configurations and $T$ be the set of transitions. Then an execution $\execution$ is an element of the set $\mathfrak{C} \times (T \times \mathfrak{C})^{*}$, such that every transition is valid ($c_n \leadsto^{t_{n+1}}_{sem} c_{n+1} )$. 
We assume all executions terminate and reach a final configuration
$\angles{\sskip, \sigma_f}$ for some $\sigma_f$.
A transition label $\transition$ has two parts
\[
\{L_n, R_n \mid n \in \mathbb{N}\}^* \;\;\times\;\; (\{ \epsilon \} \cup (Var \times Val))
\]
The first component is a \emph{fragment label}, uniquely identifying some \ttcommute{} block's fragment.  The label refers to the sequence of Proj-Left and Proj-Right rules used in the transition, with the subscript referring to the number of sequentially composed \texttt{commute} blocks. In parallel executions, the fragment label corresponds with a unique executing thread, and we may use the two words interchangeably. As an example, if there are no sequentially composed \texttt{commute} blocks, applying Proj-Left twice then Proj-Right once may yield a fragment label of $L_0L_0R_0$.
The second part of a label is either $\epsilon$ or $v \mapsto c$. The small steps only at most modify the value of one state element at a time, so this is sufficient to capture all effects.

We write $\mathfrak{C} \Downarrow_{sem} \execution$ if the initial configuration of $\execution$ is $\mathfrak{C}$, and let $\execution_f$ notate the final configuration of the execution $\execution$.


\section{A Condition for Safe Parallelization}
\label{sec:scoped-serializability}

As discussed in Sec.~\ref{subsec:overview-semantics}, our programming model is distinct from, say, transactional memory in which users explicitly fork threads.
Our goal here is instead to safely interchange the parallel semantics for the sequential semantics, \ie, show that $\sem{s}_{par} = \sem{s}_{seq}$. 
We now define this notion of safe parallelization and then introduce a new correctness condition---\emph{scoped serializability}---which is distinct and slightly stronger than serializability and show that, when combined with valid commutativity conditions, it achieves the goal.  

\subsection{Definitions and Properties}

We choose to work with semantics based on \emph{post state equivalence} rather than traces (projecting actions out of executions) because, although the latter works well for serializability, the former is  simpler in our setting. Our notion of post-state equivalence also permits weaker notions of state equivalence (\ie~\emph{observational} equivalence), and corresponds more closely to standard definitions of commutativity~\cite{KR:PLDI11,DBLP:conf/pldi/DimitrovRVK14,Bansal2018,Pirlea2021}. 

Toward defining the goal, we begin with some definitions:

\begin{definition}[Big-Step Semantics]
	\label{sem}
	For  program $s$, let $\sem{s}_{sem} : \Sigma \rightarrow \mathbb{P}(\Sigma)$ be defined as:
	\[  \sem{s}_{sem} \;\equiv\; \lambda\ \sigma_0. \;\{ \sigma_f \mid  \exists \execution: \angles{s, \sigma_0} \Downarrow_{sem} \execution \wedge \execution_f.st = \sigma_f \} \]
\end{definition}
That is, the set of final states of executions.
We say $\sem{s_0}_{sem_0}(\sigma_{0_0}) = \sem{s_1}_{sem_1}(\sigma_{0_1})$ when they are equal as sets. We say for statements $s_0, s_1$, that $\sem{s_0}_{sem_0}$ and $\sem{s_1}_{sem_1}$ are \emph{equivalent} when, $\forall \sigma, \sem{s_0}_{sem_0}(\sigma) = \sem{s_1}_{sem_1}(\sigma)$
%
%
We say $s$ is \emph{deterministic} in $sem$ when $\forall \sigma: \sem{s}_{sem}(\sigma)$ is singleton.


We say that $s$ is \emph{parallelizable} if $\sem{s}_{seq} = \sem{s}_{par}$, \ie, defined in terms of end states of a program under different semantics.
In the next subsection we will discuss a condition that ensures this goal.



There are a few key properties of the semantics, given below, that are later used in proving other lemmas or the main theorem. These properties are proved in Apx.~\ref{apx:properties}.
\begin{enumerate*}
\item Redex Uniqueness: $\forall s: \exists! H, r: s = H[r]$.
\item Conditional Determinism:
	$\forall s, \sigma, H, r: s = H[r] \wedge
	(\forall s_0, s_1: r \neq \texttt{commute(true)} \block{\block{s_0} \block{s_1}}) \rightarrow
	\exists s', \sigma', s'', \sigma'': (\angles{s, \sigma} \leadsto_{sem} \angles{s', \sigma'}) \wedge (\angles{s, \sigma} \leadsto_{sem} \angles{s'', \sigma''}) \rightarrow (s' = s'' \wedge \sigma' = \sigma'')$ .
\item Commutativity Correctness: If 
	every \texttt{commute} block guard in $s$ is a sufficient commutativity condition then $s$ is deterministic in $nd$.
\item Inclusion:
	$\forall s, \sigma: \sem{s}_{seq}(\sigma) \subseteq \sem{s}_{nd}(\sigma) \subseteq \sem{s}_{par}(\sigma)$.
	\item Deterministic Sequential:
If	$s$ is deterministic in $sem$ then $\sem{s}_{sem} = \sem{s}_{seq}$.
\end{enumerate*}

\subsection{Scoped Serializability}

We now define our correctness condition. We begin with a single execution:

\begin{definition}[Scoped serial execution] Execution $\execution$ is scoped serial if:
\[
\begin{array}{ll}
	\multicolumn{2}{l}{\forall p \in \{L_n, R_n \mid n \in \mathbb{N}\}^*: } \\ 
	((\forall \transition, \transition' \in \execution: & \transition.\mathit{fr} \text{ has prefix } p \cdot L_k \wedge 
	 \transition'.\mathit{fr} \text{ has prefix } p \cdot R_k \implies \transition \le_{\execution} \transition') \\
	\vee (\forall \transition, \transition' \in \execution: & \transition.\mathit{fr} \text{ has prefix } p \cdot L_k \wedge 
	 \transition'.\mathit{fr} \text{ has prefix } p \cdot R_k \implies \transition' \le_{\execution} \transition))
\end{array}
\]
\end{definition}
Above, $\transition.\mathit{fr}$ is the fragment label of $\transition$.  
The key idea here is to identify the \emph{scope} of commute fragments through labels, and then require a serializability condition for the L/R pair of the given scope.
Consider, \eg, a single \ttcommute{} block, possibly with children. For an execution to be scoped-serial, we require all of the transitions from one of the fragments to execute prior to all the transitions from its co-fragment (the other statement in the \texttt{commute}). Next, when there are nested \texttt{commute} blocks, the quantification over prefixes requires that we expect this same property to hold locally for all nested \texttt{commute} blocks. Without nesting, we recover the standard notion of serial.
We now define the following correctness condition, which requires that every execution is equivalent to another scoped serial execution:

\begin{definition}[Scoped Serializability] For execution $\execution$ such that $\mathfrak{C} \Downarrow_{par} \execution$,
	$\execution$ is \emph{scoped serializable}, wrt $\mathfrak{C} $ if and only if 
	\[ \exists \execution' : \mathfrak{C} \Downarrow_{par} \execution' \wedge \execution' \text{ is scoped serial} \wedge \execution_f.st = \execution'_f.st.\]
\end{definition}

A program $s$ is scoped serializable (or s-serializable) if every execution of $s$ is scoped serializable.

\paragraph{S-Serializability vs Serializability}
S-Serializability is distinct from serializability. Furthermore, serializability is an insufficient condition for parallelization. Consider 
Ex.~\ref{ex:notser} from Sec.~\ref{sec:overview}.
	Suppose that the sub-statement \ttt{x = x * 2}  locks x so that it appears to execute atomically (we defer  formalization of locks).
	If we merely require that each thread's own steps appear to act together, then an acceptable serial execution (omitting intermediate configurations) of this program is:
\[\begin{array}{lll}
	(\emptyset, & (\epsilon, y \mapsto 0), (\epsilon, x \mapsto 1), \\
	& (L_0R_0, x \mapsto x*2 ), & \text{\emph{Here} } x * 2= 2 \\
	& (R_0, \epsilon) & \text{\emph{Reads x's value as 2}} \\
    & ( R_0, y \mapsto 1), (L_0L_1, x \mapsto 0))
\end{array}\]
	After which we have $\sigma_f = \{ x \mapsto 0, y \mapsto 1 \}$. However, no pairwise re-ordering of the commutative blocks can result in such a state; if statement $\ttt{if (x>2) \{ y=1; \}}$ executes first then \tty\ will be 0, and if it executes second, then \tty\ will still be 0.
	
To adapt the traditional notion to our setting, one would permit interleavings of outer \ttcommute{} statements that could interfere with inner \ttcommute{} statements. The property (see Apx.~\ref{apx:adaptedser}) would specify order requirements on interactions between fragments that do not share the same prefix.
Naturally, our notion of S-Serializability implies this adapted notion of serializability.

\subsection{Parallelizability}

As our setting allows for a nested structure of \ttcommute{} statements, and also sequentially composed \ttcommute{} statements, these must be reflected in the parallel semantics. To ensure parallelizability, this structure must be preserved. While serializability is insufficient, we prove that scoped serializability is a sufficient condition.
\begin{restatable}{lemma}{mainlemma}
	\label{RSer and ND}
	S-Serializable($s$) $\implies \sem{s}_{nd} = \sem{s}_{par}$
\end{restatable}
\noindent
By adding Lemma~\ref{Commutativity Correctness}, we conclude our main theorem:

\begin{restatable}[Sufficient Condition for Parallelizability]{theorem}{maintheorem}
	\label{Main Theorem} If every commutativity condition in $s$ is valid  and S-Serializable($s$), then $s$ is parallelizable.
\end{restatable}
\begin{proof}
	By induction on maximum fragment length of transitions in $par$ executions of $s$. Full proof in Apx.~\ref{apx:proof}.
\end{proof}



\newcommand\conflictIns{\mathcal{C}\mathcal{I}}
\subsection{Enforcing Scoped Serializability}
\label{subsec:enforcing}

Note that any sufficient condition for pairwise serializability admits a condition for s-serializability by inductively applying it on each \ttcommute{} statement.
Imagine that we have first enforced s-serializability for a sub-configuration.
Then we can use a strategy for enforcing pairwise serializability as an inductive step for enforcing s-serializability in outer scopes.
We thus now discuss techniques for enforcing pairwise serializability (and thus s-serializability).



\paragraph{Pattern 0: Na\"{i}ve Locking}

To our knowledge, prior works have not focused on pairwise serializability, perhaps because it hasn't been useful in other settings (\eg, parallel composition). However, we may still take the na\"{i}ve approach of locking from the first read/write on a common variable to the last. For only two blocks, this likely would not yield performance gains, as it leads to mutual exclusion, but in
our nested \ttcommute{} setting, we can still aim for above 2$\times$ speedup due to the nesting of \ttcommute{} statements.

Indeed because of the structure of our setting we can \emph{use and improve} the possible gains from pairwise serializability.

\paragraph{Pattern 1: Write/ReadOnly Intersection}
\label{ssec:wrro}
In some cases, although both threads are reading/writing multiple variables, it may be that, among the \emph{commonly} accessed variables, one thread is only reading from that set.

Let $wr(s)$ be the set of variables that $s$ writes to, and $rd(s)$ be the set of variables that $s$ reads from. In the case of an instance of a builtin ADT, we include that builtin's variable name 
(accounting for aliasing)
in both $wr(s)$ and $rd(s)$ if a method is invoked on it. Define the set of conflicting variables be:
\[\begin{array}{lll}
	con(s_0, s_1) &\equiv& [wr(s_0) \cap wr(s_1)] \cup [wr(s_0) \cap rd(s_1)] 
	 \cup [rd(s_0) \cap wr(s_1)]
\end{array}\]

Now let us consider the case where that $(wr(s_0) \cup rd(s_0)) \cap wr(s_1) = \emptyset$. Then $con(s_0,s_1) = wr(s_0) \cap rd(s_1)$. We may consider the symmetrical case (with $s_0$ and $s_1$ swapped) as well.
We propose a simple program transformation can be used to enforce s-serializability in cases where this pattern holds.
For each (conflicting) variable $v \in con(s_0, s_1)$,
\begin{enumerate*}
\item Replace all reads of $v$ in $s_1$ to that of a new var $v_0$.
\item Before the \ttt{commute} block, add a statement $v_0 = v$.
\end{enumerate*}
Then all of $s_1$'s reads will refer to the state prior to execution of $s_0$, regardless of whether $s_0$ executes first or not. So every interleaving will be as if it were the case $s_1; s_0$. One can think of the above process as snapshotting \label{WrRO}the state prior to potential modifications that may cause issues when interleaved.

\paragraph{Pattern 2: Narrowing mutual exclusion}
\label{ssec:mutex sync}

Now we consider the more general case, when there are variables written by both threads. We can still use a single lock to enforce scoped serializability.
(For a formal treatment of locking in our semantics, refer to Apx.~\ref{apx:locking}.) If the threads lock on a common lock, they necessarily mutually exclude each other (and thus have no gains from parallelization), so the goal is to merely minimize the amount of time that the lock is held. Intuitively, this means we want all of the operations on conflicting variables as temporally close as possible in each thread.
As is done elsewhere~\cite{Cerny2013}, this can be done via program transformations.

We describe a method for reordering statements in commutative blocks so that each block may be treated as an atomic point with respect to one another, \ie~the point when it acquires the singular lock.
Suppose we are provided  a data-flow graph for $s_0$ and $s_1$. Identify all instructions (nodes) that refer to the set of conflicting variables $con(s_0, s_1)$. Call this set $\conflictIns{}_{s_0}$ (resp. $\conflictIns{}_{s_1}$). Next, reorder instructions to be in the following form:
\begin{enumerate}
	\item \label{Ancestor} Instructions that are only ancestors of nodes in $\conflictIns{}_{s_0}$ (\ie, instructions whose data only flows into the conflicting variables);
	\item $lock();$
	\item \label{Spine} Instructions that are in, or that are both ancestors and descendants of (possibly distinct) nodes in, $\conflictIns{}_{s_0}$.
	\item Snapshot $con(s_0, s_1);$
	\item $unlock();$
	\item \label{Descendant} Instructions that are only descendants of nodes in $\conflictIns{}_{s_0}$ (\ie, instructions who are only dependent on the conflicting variables).
	\item \label{Independent} Instructions that are neither ancestors nor descendants of nodes in $\conflictIns{}_{s_0}$.
\end{enumerate}
We use the same notion of snapshotting as in Pattern 1. This transformation is correct (assuming proper instruction ordering within each of the four groups), as it preserves all data dependencies, and it enforces pairwise serializability, as we may treat the fragments as executing sequentially in the order of reaching the lock. Indeed, we can re-order most transitions in an execution, as:
\begin{itemize}
	\item The statements in (\ref{Ancestor}) for $s_0$ can not have a dependency to the statements in (\ref{Spine}) for $s_1$, as otherwise they would cause writes in $s_0$ that are read in $s_1$, a contradiction as then they would be in $\conflictIns{}_{s_0}$, and thus in (\ref{Spine}).
	\item The statements in (\ref{Descendant}) for $s_0$ can not have dependencies from the statements in (\ref{Spine}) for $s_1$, as they have been modified to refer only to the local snapshot.
	\item All data dependencies across the blocks must involve a statement from (\ref{Spine}). So, for example, (\ref{Ancestor}) for $s_1$ does not depend on (\ref{Descendant}) for $s_0$.
\end{itemize}
Thus we can reorder all transitions in (\ref{Ancestor}) in the second thread to after (\ref{Spine}) of the first, and all statements in (\ref{Descendant}) and (\ref{Independent}) of the first thread to before the (\ref{Spine}) of the second, obtaining a serial execution.

For load balancing reasons, it may be practical to put (\ref{Independent}) at the beginning of the block when reordering $s_1$. If fact, even when locking is non-trivial, using this technique can ensure correctness at minimal performance cost with proper load balancing.

\newcommand\Xs{\bar{x}}
\newcommand\Ys{\bar{y}}
\newcommand\constr{\psi_{constr}}
\newcommand\bconstr{\overline{\psi}_{constr}}
\newcommand\smt[1]{\underline{#1}}
\newcommand\Tr{\mathit{T\!r}}
\newcommand\TrE{\mathit{T\!r}_E}
\newcommand\TrS{\mathit{T\!r}_S}
\newcommand\smtLet{\underline{\mathit{let}}}
\newcommand\smtIn{\underline{\mathit{in}}}
\newcommand\smtIte{\underline{\mathit{ite}}}
\newcommand\smtTrue{\underline{\mathit{true}}}
\newcommand\smtFalse{\underline{\mathit{false}}}
\newcommand\smtExists{\underline{\mathit{exists}}}

\newcommand\plet{{\bf\sffamily let}}
\newcommand\pin{{\bf\sffamily in}}
\newcommand\pprec{{\bf\sffamily rec}}
\newcommand\pmatch{{\bf\sffamily match}}
\newcommand\pwith{{\bf\sffamily with}}
\newcommand\ptype{{\bf\sffamily type}}

\section{Verification and Inference of Commutativity Conditions}
\label{sec:verifyinfer}

For the parallelized version of the program to be correct, \ttcommute{} conditions must be sound.
Although there are some recent techniques for verifying
or inferring commutativity conditions, these techniques do not quite fit our setting, as discussed in Sec.~\ref{sec:relwork}. For example, the abstraction-refinement technique of \citet{Bansal2018} applies to logically specified abstract data-types (ADTs) and not directly to code.
In this section we will describe an embedding of programs with \ttcommute{} statements (which may program context, involve nested \ttcommute{} statements, builtins hashtables, loops, etc.) into such logical specifications, thus enabling us to use such abstraction-refinement to synthesize conditions that can be used in our \ttcommute{} blocks, or to verify manually provided conditions.
%


Since we target the abstraction-refinement synthesis algorithm of 
\citet{Bansal2018}, we briefly recall the input to this algorithm, which is a logically specified ADT $\mathcal{O}$, defined as:
\begin{definition}[Logical ADT specification]
\[\mathcal{O} = \left\{\begin{array}{lllllllll}
 & \textsf{state} &:& (Var,Type) \textsf{list};\;\;\;\;\;& \textsf{methods} &:& \textsf{Meth list};\\ 
	& \textsf{eq}    &\subseteq& \textsf{state} \times \textsf{state}; & \textsf{spec} &:& \textsf{Meth} \rightarrow (P,Q) & 
\end{array}\right\}\]
\end{definition}

\noindent
These objects include state defined as a list of variables, an equality relation on states, a finite list of method \emph{signatures} (without implementations), and logical specifications for each method. 
The \emph{commutativity condition synthesis problem} is then, for a given pair of method signatures
$m(\Xs) : \textsf{Meth}$ and $n(\Ys): \textsf{Meth}$,
to find a logical condition $\varphi$ in terms of the object's state, and the arguments $\Xs$ and $\Ys$, such that $m$ and $n$ commute according to the method specifications from every state satisfying $\varphi$. \citet{Bansal2018} give a method for doing so based on abstraction-refinement.

\subsection{Embedding \ttcommute{} blocks as logical ADT specifications.}

The challenge we now address is how to 
translate programs with \ttcommute{} blocks into logical object specifications.


%

For a given \ttcommute{} block with statements $s_1$ and $s_2$ and their program context $E$, 
our translation $\Tr$ is defined below
and yields an object $\mathcal{O}$ with two methods $m_{s_1}$ and $m_{s_2}$.


\begin{definition}[Embedding $\Tr$] For \ttcommute\ fragment statements $s_1$ and $s_2$ in program context $E$, $\Tr(E,\ttcommute \; s_1 \; s_2) \equiv \mathcal{O}$ where 
\begin{itemize*}
	\item $\underline{\mathcal{O}.\textsf{state}}$: We create an object variable for every global or local variable in environment $E$ (that is visible to the scope containing $s_1$ and $s_2$).
	\item $\underline{\mathcal{O}.\textsf{eq}}$: 
	$
	\lambda o_1,o_2.
	\bigwedge_{v\in(\textsf{map fst } \mathcal{O}.\textsf{state})}
	o_1.v = o_2.v
	$.
	\item $\underline{\mathcal{O}.\textsf{methods}}$:
	$[ m_{s_1} : \textrm{unit} \rightarrow \textrm{unit}, m_{s_2} : \textrm{unit} \rightarrow \textrm{unit} ]$
	\item $\underline{\mathcal{O}.\textsf{spec}}$:
	$
	[ m_{s_1} \mapsto (\textsf{true},\textsf{specOf}(s_1,E)),
	m_{s_2} \mapsto (\textsf{true},\textsf{specOf}(s_2,E)) ]
	$ with $ \textsf{specOf}$ in Fig.~\ref{fig:tr}.
\end{itemize*}
\end{definition}
Intuitively the idea is to embed the current program context (including variables in scope and builtins) 
as $\mathcal{O}$'s state
and  $m_{s_1}()$ and $m_{s_2}()$ as nullary methods corresponding to $s_1$ and $s_2$, respectively,
whose pre/post specifications describe the changes to that context.

The last \textsf{spec} component translates \ttcommute{} fragment source code to a logical method specification. The overall shape of the translation is akin to verification condition generation, introducing indexed let-binding $v_i$ each time program variable $v$ is mutated and, ultimately, constraining the value of $v$ in the postcondition (with respect to the 0th bindings) in terms of the most recent let-binding.
(Programs in our language are total and thus, preconditions are simply \textsf{true}.)
The output of our translation is a logical  formula, written in SMTLIBv2 (see Section 5.2 of \citet{smtlibpdf}):
\newcommand\smtTerm{\gamma}
\[\begin{array}{llll}
	& \varphi &::=&  \varphi \vee \varphi \mid \neg \varphi \mid \smtIte\; \varphi \; \varphi \; \varphi \mid \smt{f} \; \bar{\varphi} \mid \smtLet\; v \; \varphi \; \smtIn \; \varphi \mid \smtExists \; v \; \smtIn \; \varphi \mid \varphi \; \underline{\otimes}\; \varphi \mid \smtTrue \\
\end{array}\]
Greek letters are used for logical terms. $\Phi$ will be used for the type of terms.
A formula $\varphi$ is a well-sorted term with sort boolean. The terms we use include $\smtLet$, $\smtIte$, and $\smtExists$, underlining formulae constructors to distinguish from our language and algorithm.

\newcommand\doubleplus{\text{+\kern-0.5ex+\kern0.5ex}}

\begin{figure}
\begin{tabular}{|l|l|}
\hline
\begin{minipage}[t]{2.65in}
\begin{program}[style=sf]
\ptype{} idmap = $Var \rightarrow \mathbb{N}$\\
   $\;$\\
   \plet\tab\ \pprec{} $\TrE$ ($e$:expr) (id:idmap) : \\
   	  idmap $\times \Phi \times$ (id $\times \Phi) list$ = \\
      \pmatch{} $e$ \pwith{}\\
      $\mid$ $v$ $\rightarrow$ (id, $\smt{v_{(\textsf{id v})}}, []$)\\
      $\mid$ $e' \otimes$ \tab\ $e''$ $\rightarrow$ 
            let ($\varphi_{L}$, binds$_{L}$) = $\TrE$ $e'$ id in\\
            let ($\varphi_{R}$, binds$_{R}$) = $\TrE$ $e''$ id' in\\
            (id, $\varphi_{L} \underline{\otimes} \varphi_{R}$, binds$_{L}$ \doubleplus binds$_{R}$)\untab \\
      \ldots\\
      $\mid$ b\tab uiltin adt $v_{adt}$ m $\bar{e}$ $\rightarrow$ \\
         \plet{} \tab (id\_{ret} :: $\overline{\varphi_{args}},\overline{\text{binds}}$) = \\
         	map ($\TrE\; \bullet\;$ id) $\bar{e}$ \pin{} \untab \\
         \plet{} adt\_id = id $v_{adt}$ \pin{}\\
         \plet{} \tab ($\varphi$, spec\_binds) = \\
            \boxed{\textsf{spec adt $v_{adt}$ m $\bar{\varphi_{args}}$ adt\_id}} \pin{} \untab \\
         (id[id$_\text{ret}$ $mapsto$ id[id$_\text{ret}$] + 1], \\
         $\varphi$, $\overline{\text{binds}}$ \doubleplus spec\_binds)\\
   \untab
   \untab
   $\;$  \\
   \plet{}\tab\ specOf ($s$ :stmt) : $\Phi$ = \\
   \plet{} $s_{flat}$ = flatten $s$ \pin{}\\
   \plet{} $s'$ = summarizeLoops($s_{flat}$) \pin{}\\
   \plet{} id0 = $[ v \mapsto 0 \mid v \in Vars(s)]$ \pin
\\
   \plet{} inits = \{ ($v^0_0$ $v^0$) ($v^1_0$ $v^1$) ... $\mid v^i \in Var(s)$\} \pin{}\\
   \plet{} (\_, $\varphi$) = ($\TrS$ $s_{flat}$ id0) \pin $\;$
   ($\smtLet$\tab\ \; inits \; $\smtIn$ \; $\varphi$)\\
\end{program}

\end{minipage}
      &
\begin{minipage}[t]{2.53in}
\begin{program}[style=sf]
   \plet\tab\  \pprec{} $\TrS$ ($stmts$:stmt list) (id : idmap) : $\Phi$ =\\
   \pmatch{} $s$ \pwith\\
   $\mid$ [] $\rightarrow$ $\bigwedge_{v\in\textsf{dom(id)}}. v_{new}$ = $v_{(\textsf{id v})}$ \\
   $\mid$ ($v$ \tab = $e$) :: tl $\rightarrow$
     \plet{} (id', $\varphi_e$, bindings) = ($\TrE$ $e$ id) \pin{}\\
       $\; \smtLet{}$ \tab $\smt{bindings}$, $\smt{v_{(\textsf{id' v}+1)}}$ $\varphi_e$ $\smtIn{}$ \\
       $\TrS$ tl (id'[$v$ $\mapsto$ (id' $v$)+1])\untab\untab\\
   
   $\mid$ (if\tab\ $e$ then $s_1$ else $s_2$) :: tl $\rightarrow$ \\
      \plet{} (id', $\varphi_e$, bindings) = ($\TrE$ $e$ id) \pin \\
      $\smtLet{}$ \tab $\smt{bindings}$ $\smtIn{}$ \\
        ($\smtIte$ $\varphi_e$ \tab ($\TrS$ (s$_1$ \doubleplus tl) id') \\
			($\TrS$ (s$_2$ \doubleplus tl) id'))\untab\untab\untab\\
   $\mid$ (co\tab mmute $e$ $s_1$ $s_2$) :: tl $\rightarrow$\\
      \boxed{\textsf{$\TrS$ id ($s_1$\! \doubleplus \!$s_2$ \doubleplus tl)}}\untab \\
   $\mid$ (as\tab sume $e$) :: tl $ \rightarrow$ \\
      \plet{} (\_, $\varphi$, \_) = $\TrE$ $e$ id \pin{} \\
      $\varphi \wedge$ $\TrS$ tl id \untab \\

   $\mid$ (ha\tab voc $hid$) :: tl \ $\rightarrow$ \\
   	  \plet{} $hid_{havoc}$ = freshvar () \pin{} \\
      $\smtExists$ $hid_{havoc}$ $\smtIn{}$ \\
      $\; \smtLet{}$ $\smt{hid_{(\textsf{id hid}+1)}}$ $hid_{havoc}$  $\smtIn{}$ \\
      $\TrS$ tl id[$hid$ $\mapsto$ (id $hid$)+1])
\untab\\
\end{program}
\end{minipage}\\
\hline
\end{tabular}
\caption{\label{fig:tr} Reducing a \ttcommute{} fragment's statement $s$ to a logical method specification.}
\end{figure}


The translation algorithm \texttt{specOf} is given in Fig.~\ref{fig:tr}. 
First we replace all sequential composition of statements with lists of statements, where $\sskip$ is the empty list.
%
We proceed in an SSA-like manner,
creating an initial mapping \textsf{id0} of indices for each variable and construct a $\smtLet{}$ term binding $v_0$ with $v$ for each  variable that is defined in the context $E$ 
and accessed in the body of the \ttcommute{} fragment.
This binding will let us ultimately relate variable $v_{new}$ in terms of its initial value $v$.
We then begin with translation $\Tr_S$ on $s_{flat}$. A call to 
$(\Tr_S \; s \; \textsf{id})$ recursively constructs the translation to a logical expression representing the postcondition of the block.
Within a given call to $(\Tr_S \; s \; \textsf{id})$, if $s$ is a final \textsf{skip} statement (i.e., if s is the empty list), then an innermost formula is constructed that relates each $v_{new}$ to the most recent binding for $v$ in the ID map.
In the assignment case, $e$ is first translated, which may accumulate additional variable bindings, namely when $e$ contains a call to a builtin ADT.
Then a fresh binding for variable $v$ is created with the new value.
In the \texttt{if/else} case, 
we recursively translate the condition expression $e$ to obtain a logical condition $\varphi_e$. We then represent branching via $\smtIte$, with condition $\varphi_e$ and the remainder of the block \textsf{tl} appended to both the branches, recursively translated.
Our semantics of \ttcommute{} statements allows this translation to treat them as sequential composition in either order, so we simply choose $s_1;s_2$ and recursively translate their concatenation. 
Loops are supported by instrumenting loop summaries with \texttt{havoc} and \texttt{assume}, as detailed at the end of this subsection. 
%
As usual, \ttt{havoc} existentially quantifies a new value and \ttt{assume} introduces a new specified constraint, both with respect to the current binding IDs.

The translation of expressions $\Tr\;e\;\textsf{id}$ 
returns a triple (an updated \textsf{id} map, a translation of $e$ into logic, and possibly new variable bindings),
and is mostly straight-forward, except in the case of a builtin ADT. This case requires us to instantiate the ADT's logical specification for the current calling context, so we first recursively translate each of the arguments in $\bar{e}$, obtaining a set of local bindings $\overline{\textsf{binds}}$. We then use a helper method \textsf{spec} (omitted) to instantiate the arguments into the ADT's post-condition, and construct an SMT expression $\varphi$ describing the return value and additional bindings for the return value(s) (on a logical level, we may consider the ADT's updated state to be a return value, \eg~a hashtable has a set of keys, an integer size, and a finite map of keys to values). In this way, the specification internals are ``ferried'' through the nested $\smtLet$ statements and constraints in the final embedding.
The translation of builtins is illustrated (among other things) in the following example:

\begin{example}[Illustrating the embedding]
\begin{lstlisting}[style=vcy,language=VCY]
commute (c$>$0) {
  $s_1$: { c=c+1; } 
  $s_2$: { if (c$>$0) { tbl[c] = 5; c = c-1; }
\end{lstlisting}
\end{example}
\noindent
We label the fragments above by their statements.
We describe how \texttt{specOf} converts $s_2$ to a formula. A map \textsf{id0}=[$c \mapsto 0, b \mapsto 0, tbl \mapsto 0$] is created, as well as let-bindings $((b_0 \; b) (c_0 \; c) (tbl_0 \; tbl))$.
	The first $\TrS$ call is for the if-then-else, and it recursively translates the condition: 
	$\TrE \; \texttt{c>0} \; \textsf{id0}$, which
	takes the $c \otimes 0$ case, recursively translates 
	$c$ to $c_\textsf{id0 c}$ and then
	returns (\textsf{id0}, $(\overline{>} c_0 0)$, []).
	Now in the then branch, $\TrS$ recursively translates the sequential composition.
The assignment is syntactic sugar for \texttt{ht\_put(tbl, c, 5)}. Here the recursive translation of the first argument translates argument \ttc{} to $c_0$, and then a call is made to \textsf{spec}  \texttt{ht $tbl$ put $c_0$ 5 (id $tbl$)}. This helper function instantiates the specification to \ttt{put} as follows:
	\[\def\arraystretch{0.85}\begin{array}{ll}
		\varphi_{put} \equiv &(\smtIte \; (mem \; c_0 \; tblK_0)
		\; (\neq 5 \; (select \; tblM_0 \; c_0))\;  \smtTrue{})\\
		binds \equiv &
		\left(
		\begin{array}{lll}
			(tblK_1 (\smtIte  & (mem \; c_0 \; tblK_0)\; tblK_0 \; (ins \; c_0 \; tblK_0)) \\
			(tlbS_1 (\smtIte  & (mem \; c_0 \; tblK_0)\; tlbS_0 \; (tlbS_0 + 1)) \\
			(tblM_1 (\smtIte  & (mem \; c_0 \; tblK_0) \\
			& (\smtIte \; ((select \; tblM_0 \; c_0) = 5)\;\;\; tblM_0 \;\;\; (store \; tblM_0\; c_0 \; 5)) \\
			& (store \; tblM_0\; c_0 \; 5)))
		\end{array}\right.\\
	\end{array}
	\]
	%
	The above bindings involve three variables $tblK_1$, $tblS_1$, $tblM_1$ representing, resp., the set of keys, the size, and the finite map of the hashtable after the \textsf{get} operation, with respect to the corresponding previous variables $tblK_0,tblS_0,tblM_0$.
	These variables model the standard hashtable semantics.
	Returning to the $\TrS$ of the assignment statement, the new value of the state is bound to, then the increment of \ttc\ is translated
	to $\lambda\;k. \smtLet{} \; (c_1 (c_0 - 1)) \; \smtIn{} \; k$ by a recursive call.
	After the assignment a \textsf{skip} will be introduced, so 
	an innermost constraint of the form
	($c_{new} = c_1 \wedge tblK_{new} = tblK_1 \wedge ...$) is constructed.
	Returning to the original $Tr$ call on the if-then-else, the full translation is assembled:
	\[\begin{array}{l}
		\smtLet{} \; ((c_0 \; c) \ldots (tblK_0 \; tblK) \ldots) \smtIn{} \; \smtTrue{} \; \wedge \; \\
		\;\;\; (\smtIte{} \; (c_0 > 0) \;\\
		\;\;\;\;\;\;\;\;\;\; (\smtLet{} \; binds \; \smtIn{} \; (\smtLet{} \; (b_1 \; \varphi_{put}) \; \smtIn{} \\
		\;\;\;\;\;\;\;\;\;\;\;\;\;\;    (\smtLet{} \; (c_1 (c_0 - 1)) \; \smtIn{} \\
		\;\;\;\;\;\;\;\;\;\;\;\;\;\;\;\;\;\;\; (c_{new} = c_1 \wedge tblK_{new} = tblK_1 \wedge \ldots ))
	\end{array}\]


\emph{Loop summaries.} Loops are supported through summaries that can be computed by other techniques (\eg~\cite{Ernst2020,Kroening2008,Xie2017,Silverman2019}). We calculated loop summaries (see Apx.~\ref{apx:loopsum}) using the Korn tool. 
In some cases, Korn was not precise enough, so we manually strengthened the summaries, but we believe this work could be fully automated, and that it is orthogonal to our main contributions.
We then replace loops in \ttcommute{} blocks with their summaries using \texttt{havoc} and \texttt{assume} statements.
Note that in general, approximation of loops may not be precise enough to ensure commutativity~\cite{Koskinen2021}, however,
this is not a problem in our setting because our commute conditions are verified/inferred with the requirement that post-states be exactly equal.
Thus, we would fail to verify/infer any conditions in the presence of insufficiently precise loop summaries.




\subsection{Verifying and Inferring \ttcommute{} conditions}

Our embedding allows us to use SMT tools to verify \ttcommute{} conditions and the abstraction-refinement algorithm of \citet{Bansal2018} to synthesize them.

\emph{Verification.} To verify a provided commutativity condition $\varphi$, we construct a series of constraints, modeling the
execution of $o.m_{s_1}$ and $o.m_{s_2}$ in either order
denoted $o.m_{s_1} \bowtie o.m_{s_2}$ and assert the
negation of $\varphi \implies o.m_{s_1} \bowtie o.m_{s_2}$.
%
The correctness of our technique is based on the soundness of our embedding, as described in the following theorem:
\begin{theorem}[Embedding soundness]\label{thm:embed} For every \ttcommute{} $s_1$ $s_2$,
	let $o = Tr(E,\ttcommute \; s_1 \; s_2)$. Then
	if valid($\varphi \implies o.m_{s_1} \bowtie o.m_{s_2}$) = $\varphi$, then
	$\varphi$ is a valid commutativity condition for
	\ttcommute{} $s_1$ $s_2$.
\end{theorem}



\emph{Inference.}
Often it is even more convenient for commute conditions to be automatically synthesized.
To this end our embedding also allows us to use abstraction-refinement~\cite{Bansal2018}, which takes as input the kinds of objects that we construct.
As discussed in Sec.~\ref{sec:impl} we re-implemented this algorithm, adding some improvements to that algorithm (\eg~in the predicate selection process) that are beyond the scope of this paper. 
The resulting commutativity conditions we infer are sound, again due to the soundness of our embedding (Thm.~\ref{thm:embed}).

\emph{Completeness.}
Commute conditions are akin to Hoare logic preconditions and, as such, they are subject to expressibility concerns of the assertion language. However, commute conditions need not be complete for two reasons: (i) as opposed to preconditions, when commute conditions don't hold we default to sequential execution, rather than undefined behavior and (ii) consequently, it is always fine to over-approximate commute conditions.
As we show in Sec.~\ref{sec:eval}, in our \benchCount{} benchmarks, there were not any cases where either a precise commute condition or a sensible sound approximation could be used.

%
%


\section{Implementation: Veracity}
\label{sec:impl}

Our implementation comprises an interpreter capable of parallel execution (and sequential execution for testing), as well as analyses for verification and synthesis of commute conditions (Sec~\ref{sec:verifyinfer}).
For simplicity we focused on building a concurrent interpreter, deferring backend compilation matters to future work.

\emph{Interpreter.}
Threads are implemented with Multicore OCaml, as OCaml does not natively support true multi-threading. Considerations are made in the interpreter to minimize shared memory between threads.
When a \ttcommute{} statement is reached and its guard evaluates to \texttt{true}, threads are spawned for each parallel block. Using thread pools did not substantially improve performance over spawning as-needed. Each thread is given a copy of the global environment and call stack, permitting thread-local method calls. Sibling threads' environments will only share references to variables in shared scope, namely global variables, arguments of the current method, and variables defined within the method in scope of the current block.
When a commute statement's guard evaluates to \texttt{false}, the blocks are executed in sequence. The Veracity executable includes a flag to force all commute guards to evaluate to \texttt{false}, allowing us to compare parallel-vs-sequential execution times.


A key benefit of \ttcommute{} blocks is that hashtable operations can be implemented with a linearizable hashtable (\eg~\cite{Purcell2005,Liu2014}) in the concurrent semantics. To confirm the benefit, we built a foreign function interface to the high-performance NSDI'13/EuroSys'14 concurrent hashtable~\cite{Fan2013,Li2014} called \ttt{libcuckoo}~\cite{libcuckoo}. 

There are minor differences and syntactic sugar between the formal semantics and the implemented Veracity language. (See Apx.~\ref{apx:sugaring}.) 
Conversely, currently our theory permits only exactly two fragments per \ttcommute{} statement, although our implementation is unbounded.

\emph{Verifying and inferring \ttcommute{} conditions.}
We implemented the embedding described in Sec.~\ref{sec:verifyinfer}.
This translation then generates a specification for an object $\mathcal{O}$, with \ttcommute{} blocks embedded as logical specifications, given as OCaml SMTLIBv2 expressions.
For verification and abstraction-refinement inference, we implemented a new version of Servois~\cite{servois} as an OCaml library/module for a few reasons:
\begin{itemize*}
\item Improve performance and closer integration. $\mathcal{O}$ is now an OCaml type, constructed by Veracity's compiler and passed to our new abstraction-refinement  library. 
\item Support more/other SMT solvers. We used CVC5~\cite{cvc5} for all of our experiments, but can also use CVC4, Z3, and plug in other solvers.
\item Improve/tune the predicate generation and predicate choice techniques.
\item Add support for verifying user-provided \ttt{commute} conditions.
\end{itemize*}

\noindent
To improve abstraction-refinement we also provide possible terms from which candidate predicates are constructed. We extract these terms from the syntax of the Veracity program, as well as practically selected constants which helped reduce the size of the synthesized conditions. We also extract terms from the builtin ADT specifications.
Finally, after \ttcommute{} conditions are synthesized, we then reverse translate back to Veracity expressions.



\emph{We will publicly release our implementation of all of the above.}

\section{Evaluation}
\label{sec:eval} 


\newcommand*\benchcplx{Cplx}
\newcommand*\benchtablerowstretch{0.9}
\newcommand*\benchtabletabcolsep{0.2em}
\newcommand*\benchcorrect{\ding{51}}
\newcommand*\benchincorrect{\ding{55}}
\newcommand*\benchunknown{
	{\fontfamily{phv}\selectfont ?}
}
\newcommand*\benchfailure{Failure}
\newcommand*\benchna{\textemdash}
\newcommand*\benchtimeout{\tiny\StopWatchEnd}

We evaluated our work against four goals:
\begin{enumerate}
\item Determine whether \ttcommute{} conditions could be automatically generated (Sec.~\ref{sec:verifyinfer});
\item Determine whether scoped serializability can be enforced 
using our locking patterns (Sec.~\ref{subsec:enforcing});
\item Confirm the common-sense expectation that speedups can be seen when the duration of sufficiently independent \ttcommute{} fragments increase; and
\item Whether, despite introducing a new programming paradigm, existing applications could be adapted and exploit \ttcommute{} statements.
\end{enumerate}

\emph{Experimental setup.}
All experiments below were run on a machine with a machine with an
AMD EPYC 7452 32-Core CPU, 125GB RAM, Ubuntu 20.04, and Multicore OCaml 4.12.0

\begin{figure}
\footnotesize
\begin{center}
{\bf Correctness of Commute Conditions}
\end{center}

\noindent
{\renewcommand{\arraystretch}{\benchtablerowstretch}\setlength{\tabcolsep}{\benchtabletabcolsep}\footnotesize
\begin{tabularx}{\columnwidth}{|l|r|X|}
	\hline
	\multicolumn{3}{|l|}{
	{\bf Group 1: Automatically Inferred Commute Conditions.}
	All benchmarks, except those below in group (3).
	}\\
	Program & Time (s) & Inferred Conditions \\
	\hline
	\input{inference_table}
	\\ \hline
\end{tabularx}}




{\renewcommand{\arraystretch}{\benchtablerowstretch}\setlength{\tabcolsep}{\benchtabletabcolsep}\footnotesize
\begin{tabularx}{\columnwidth}{|l|r|r|r|X|}
\hline
\multicolumn{5}{|l|}{
{\bf Group 2: Automatically Verified Commute Conditions.}
Benchmarks for which inference output was suboptimal.
}\\
	Program & Time (s) & Verified? & Complete? & Provided Condition \\
	\hline
\input{verify_table}
	\\\hline
\end{tabularx}}

{\renewcommand{\arraystretch}{\benchtablerowstretch}\setlength{\tabcolsep}{\benchtabletabcolsep}\footnotesize
\begin{tabularx}{\columnwidth}{|l|X|}
\hline
\multicolumn{2}{|l|}{
{\bf Group 3: Unverified Examples.} Including case studies.}\\
Program & Manual commute condition  \\
\hline
	\texttt{crowdfund} & \ttt{true} \\
	\texttt{dihedral} & \texttt{(!s1 \&\& !s2) || (s1 \&\& !s2 \&\& (r2 == 0 || (n \% 2 == 0 \&\& r2 == n / 2))) || (!s1 \&\& s2 \&\& (r1 == 0 || (n \% 2 == 0 \&\& r1 == n / 2))) ||  (s1 \&\& s2 \&\& r1 == r2) } \\
	\texttt{filesystem} & \texttt{fname1 != fname2} \\
	\texttt{ht-fizz-buzz} & \texttt{true; true} \\
\hline
\end{tabularx}}

\caption{\label{fig:inference} Inference and/or verification of \ttcommute{} blocks. We first attempted to infer commute conditions for all programs (Group 1), except for those with features like strings, side-effects, etc. (Group 3). When the inferred condition in Group 1 was overly complex or reflected only trivial executions, we then manually provided a condition and attempted to verify it (Group 2). }


\end{figure}

\paragraph{Benchmarks.}
We created a total of \benchCount\ example programs 
with \ttcommute{} blocks
that
use a variety of program features including 
linear arithmetic, nonlinear arithmetic, 
arrays, builtin hashtables, nested \ttcommute{} blocks,
loops and some procedures.
The complete set of benchmarks are given in Apx.~\ref{apx:benchmarks} and in the supplemental materials. 

\paragraph{Inferring/verifying \ttcommute{} conditions.}
Fig.~\ref{fig:inference} shows the results of inference/verification.
We first applied our procedure to infer commute conditions for all benchmarks (Group 1), except for \texttt{crowdfund}, \texttt{dihedral}, \texttt{filesystem} and \texttt{ht-fizz-buzz}, which have complicated loops or interprocedural calls that we could not translate to logic. We manually provided conditions for these cases, as seen in Group 3 in Fig.~\ref{fig:inference}.
Those for which we were able to infer a condition are listed in Group 1 in Fig.~\ref{fig:inference}, along with the time taken to infer and the resulting condition.
Inference was set to time out at 120 seconds. The benchmarks for which inference timed out are marked with a clock.
In \ttt{nested} examples, there are multiple \ttcommute{} blocks, so we list all conditions.
In most of the benchmarks we were able to infer concise and useful commute conditions in a few seconds or fractions of a second. 



In cases where inference leads to an overly complex condition or a condition that only applies to trivial states, we manually provided a better one and used our Sec.~\ref{sec:verifyinfer} embedding to verify it.
For this subset of the benchmarks,
the commute conditions and results are given in
the Group 2 of Fig.~\ref{fig:inference}.
We also report the time, the verification result,
and whether it found the provided condition to be complete.
We were able to verify or reject the condition in all cases in a fraction of a second, and in all but one case we were able to report on completeness as well. 

\begin{figure}
	\includegraphics[width=\columnwidth]{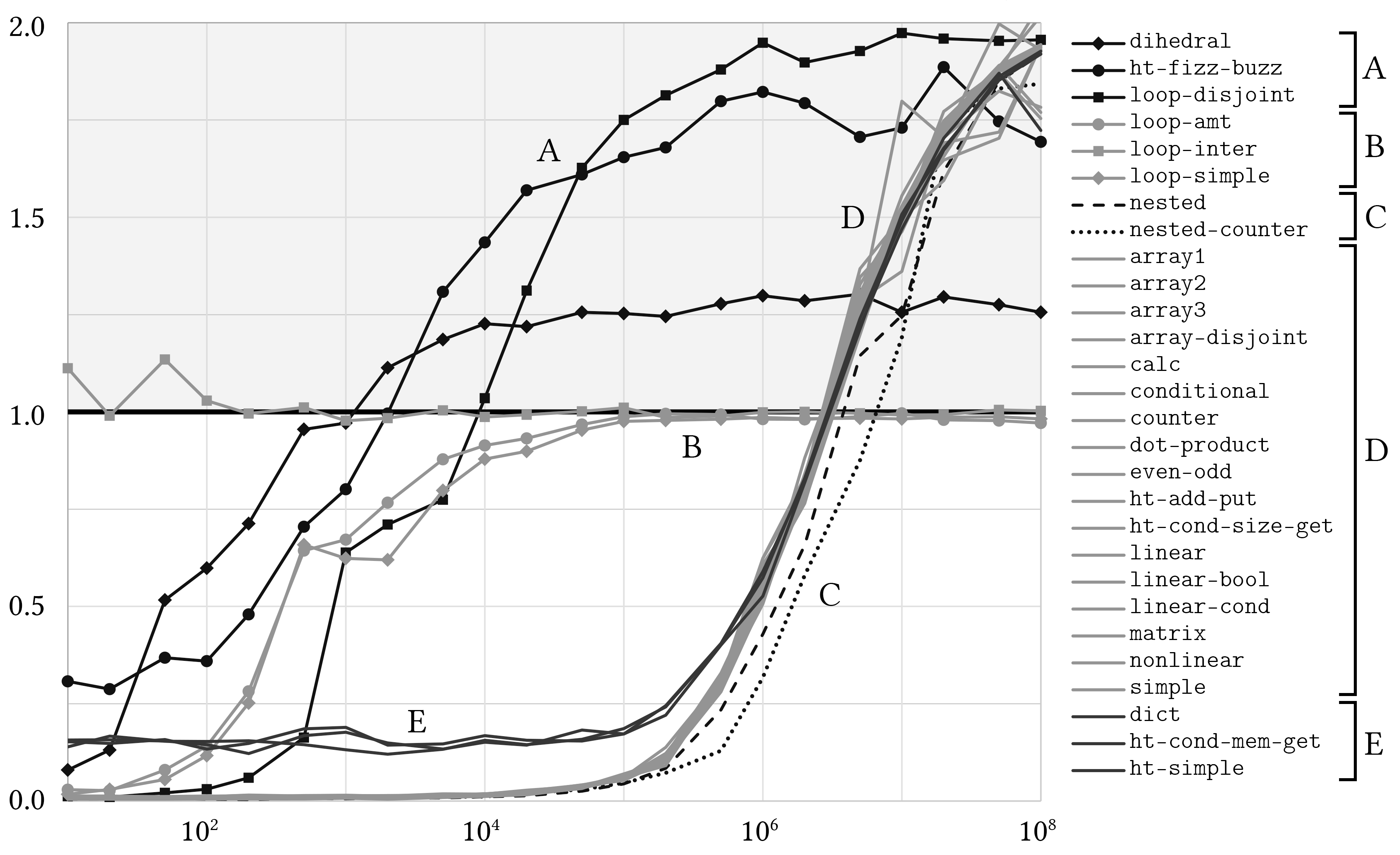}
	\caption{Parallel-to-sequential speedup for benchmark programs versus computation size. The computation size x-axis is logarithmically scaled. Each trendline is for a different program.
Many benchmarks had similar performance characteristics. For legibility of the overall plot, we color them all (Group D) in medium gray. The other group labels also indicate similar performance characteristics.}

	\label{fig:speedup}
\end{figure}

\paragraph{Enforcing scoped serializability.}
In Sec.~\ref{subsec:enforcing} we describe how to enforce 
scoped serializability. We applied these methodologies to the benchmarks, but it could be automated in the future.
For many benchmarks (\eg~\ttt{counter}), a single lock could be na\"{i}vely synthesized to protect against updates to a shared set of variables (Pattern 0).
We used our snapshot approach (Pattern 1)
on \ttt{commute1} and \ttt{simple} to avoid the need for locks at all.
Other programs (\ttt{conditional}, 
\ttt{matrix}, 
and \ttt{loop-disjoint})
did not need any locks.
For \ttt{dict}, the only conflict is a single access to an already linearizable concurrent hashtable.
\ttt{dihedral} and \ttt{ht-fizz-buzz} are serializable due to loop disjointness.
%
Note that the benchmarks displayed in Apx.~\ref{apx:benchmarks} are the original programs, to avoid confusion. For our manually applied locking, please see the
folder \ttt{benchmarks-preprocessed} in  \ttt{benchmarks.zip} from the supplemental materials.

%
%

\paragraph{Speedup.}
We next sought to confirm that, depending on the size of the computation, parallelization offers a speedup over sequential execution.
We executed the programs and collected execution time measurements.
Our benchmarks (see Apx.~\ref{apx:benchmarks}) all involve some pure computation(s) of parametric size $N$.
We varied this problem size, and recorded the speedup ratio (sequential execution time divided by parallel execution time).


We took the geometric mean of the speedup ratios across 4 trials and plotted it against problem size. The graph is shown in Fig.~\ref{fig:speedup}.
On the x-axis we increased the problem size $N$ exponentially. 
%
Many benchmarks had very similar performance. For legibility of the overall plot, we grouped and colored those with similar performance characteristics.
(The group labels do not imply any particular semantic connection between the benchmarks.)
In most of the benchmarks, speedups were observed and asymptotically approached the expected 2$\times$ mark. Most of the payoff occurred when the problem size was above 1,000,000, with some payoff occurring when the problem size was above 1,000.
In \ttt{dihedral}, the two threads can have differently sized workloads.

For the four benchmarks involving loops, we initially found no speedup. While this is expected for \ttt{loop-inter} (because it only commutes in trivial cases) and for \ttt{loop-simple} (because threads had to hold the lock for the entire loop), we discovered that the interpreter exhibited contention when concurrently accessing the shared (and pure) context. Consequently the \ttt{loop-disjoint} and \ttt{loop-amt} examples also had no speedup. This could be a subtle bug in Multicore OCaml. To confirm that this is merely a bug and that our approach should yield speedup, we edited \ttt{loop-disjoint} so that the fragments only accessed local variables (we do not do this in the original benchmark because the translation does not translate scope), and indeed we observed speedup, as plotted in Fig.~\ref{fig:speedup}.

\paragraph{Case studies.} As \ttcommute{} blocks are novel, there are no existing programs with them. Thus we explored two case studies whose concurrency could be re-formulated as Veracity programs with \ttt{commute} statements.

\begin{wrapfigure}[14]{r}[34pt]{3.9in}
	\begin{tabular}{cc}
		\begin{minipage}[b]{0.3\columnwidth}
			\begin{subfigure}{\columnwidth}
				\includegraphics[width=\columnwidth]{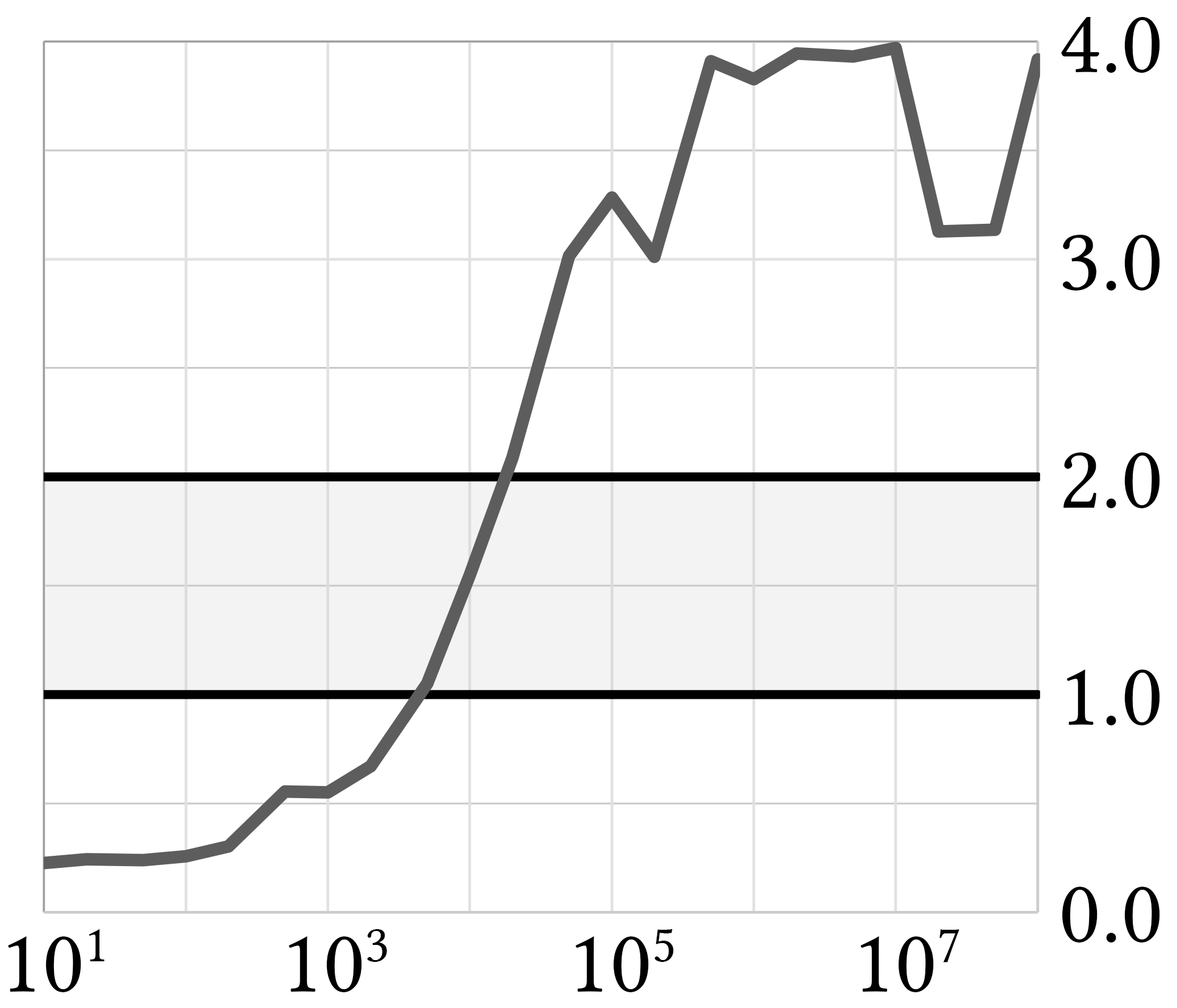}
				\subcaption{Crowdfund}
				\label{crowdfund}
			\end{subfigure}
		\end{minipage}
		\begin{minipage}[b]{0.3\columnwidth}
			\begin{subfigure}{\columnwidth}
				\includegraphics[width=\columnwidth]{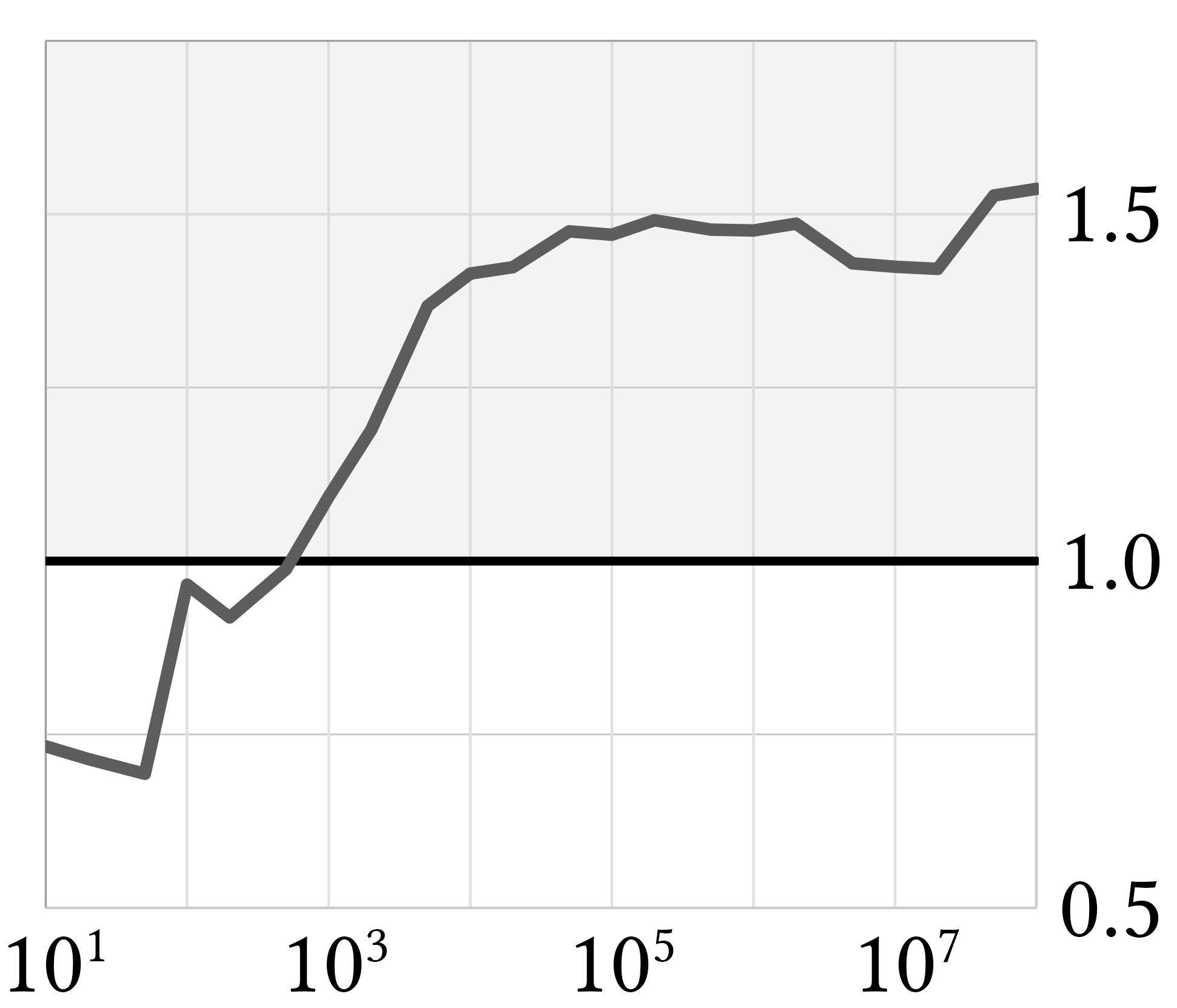}
				\subcaption{Filesystem}
				\label{filesystem}
			\end{subfigure}
		\end{minipage}
	\end{tabular}
\caption{Benchmark results for case studies.}
\end{wrapfigure}

\emph{Crowdfund Smart Contract.} Miners and validators in blockchain systems repeatedly execute enormous workloads of smart contract transactions. These transactions are currently executing sequentially and recent proposals have been made to leverage multicore architectures through parallelization~\cite{Dickerson2017,Saraph2019,Tran2021} or through sharding~\cite{Pirlea2021}.
We modeled parts of an Algorand crowdfund smart contract\footnote{\url{https://developer.algorand.org/solutions/example-crowdfunding-stateful-smart-contract-application/}} as a case study. For performance, we tested four \texttt{donate} operations in parallel. Busy waits were added at the end of each donate. As expected, we saw significant speedups; the plot of speedup against problem size can be found in Fig.~\ref{crowdfund}.
This example also illustrates that even better speedups can be achieved using \ttt{commute} blocks with more than 2 fragments.

\emph{Filesystem operations.}
A common low-conflict use case of hashtables is mapping directories to files. This could be in a user application, but also potentially in the implementation of a filesystem (\eg~ScaleFS~\cite{scalefs,Clements2015}).
We modeled filesystem operations as veracity programs. Writing to files commutes when different files are being written to.
The results are shown in Fig.~\ref{filesystem}. When writing sufficiently large files, we achieved large speedups (approaching about 1.85$\times$).

\section{Related Work}
\label{sec:relwork}

To our knowledge no prior works have explored bringing commutativity conditions \emph{into} the programming language syntax, and the ramifications thereof. We now survey other related works.

It is long known that there is a fundamental connection between commutativity and concurrency, dating back to  work from the database communities describing how to ensure atomicity in concurrent executions with locking protocols based on commutativity (\eg~\cite{weihlcommu,commu,korth}).
Compiler-based parallelization of such arithmetic operations was introduced by \citet{rinard}. 
Later commutativity \emph{conditions} have been used in many concurrent programming contexts such as optimistic parallelism in graph algorithms~\cite{Kulkarni2008}, transactional memory~\cite{ont,ppopp08,Dickerson2019}, 
dynamic analysis~\cite{DBLP:conf/pldi/DimitrovRVK14},
and blockchain smart contracts~\cite{Dickerson2017,jar21,Pirlea2021}.
Closed or open nested transactions~\cite{ont} permit syntactically nested transactions. Although nested transactions (NT) might seem akin to nested \ttcommute{} statements, NTs still fall within the realm of a programming model in which users write a concurrent program (explicitly forking and so on), as opposed to our sequential \ttcommute{} statements.

In more recent years several works have focused on \emph{reasoning} about commutativity.
\citet{DBLP:conf/cav/GehrDV15} describe a method based on black-box sampling.
\citet{aleen} and \citet{Tripp2011} identify sequences of actions that commute (via random interpretation and dynamic analysis, resp.). 
\citet{KR:PLDI11} \emph{verify} commutativity conditions from specifications. \citet{Bansal2018} synthesize commutativity conditions from ADT specifications in the tool Servois~\cite{servois} (which we build on top of in Sec.~\ref{sec:verifyinfer}). \citet{Koskinen2021} verify commutativity conditions from source code. 
\citet{Pirlea2021} discover commutativity through a static analysis based on linear types and cardinality constraints.
\citet{NGYFS2016} describe a tool for weak consistency that checks commutativity formulae. \citet{Houshmand2019} describe commutativity checking for replicated data types.

Numerous works are focused on synthesizing locks.
\citet{Flanagan2003} described a type-based approach, and
\citet{Vechev2010} used abstraction-refinement.
\citet{gulwani} and 
\citet{Golan2015} describe automatic locking techniques for ensuring atomicity of, respectively, transactions and multi-object atomic sections.



\section{Discussion and Future Work}

To our knowledge we have presented the first work on introducing \ttcommute{} statements into the programming language. We have given a formal semantics, a correctness condition, methods for ensuring parallelizability, and techniques for inferring/verifying \ttcommute{} conditions. This work is embodied in the new Veracity language and front-end compiler.

\emph{Future Work.} 
On the practical side, top priorities for future work are to explore back-end compilation strategies to emit, \eg, concurrent IR. 
Our work can also be combined with other parallelization strategies such as promises/futures~\cite{Liskov1988,Chatterjee1989}.
In our ongoing implementation we seek to integrate existing data flow analyses to reduce locking while enforcing scoped serializability.
%
%
On the theoretical side we could generalize to $N$-way \ttcommute{} statements and some notion of object-oriented encapsulation, \ie,  arbitrary concurrent objects.
With more encapsulated expert-written concurrent objects, more sequential \ttcommute{} programs can be written to exploit them.

Our work can also be combined with invariant generation, which can inform commutativity inference, as seen in the following example:
\begin{example}[Combine with static analysis]\label{ex:staticanalysis} 
\begin{lstlisting}[style=vcy,language=VCY]
|\textcolor{blue}{\{ y < 0 \}}| /* Example invariant */
commute _ {
  { y = y + 3*x; }  /* if x is negative, this will reduce y */
  { if (y<0) { x=2; } else { x=3; } } /* sensitive to whether y went below 0 */
}
\end{lstlisting}
\end{example}
\noindent
Without the static knowledge of the invariant \texttt{y $<$ 0} before the commute condition, an excessively complex commute condition would be inferred. However, if we first perform a static analysis, there are fewer choices to be made during abstraction-refinement, quickly leading to a simpler and more context-sensitive commute condition (in this case: \texttt{0 $>$ y + 3*x \&\& 2 == x}).
Our embedding already supports such context information so in the future we aim to integrate Veracity with an abstract interpreter.




\vfill
\pagebreak


\vfill
\pagebreak

\appendix
\section{Redex reduction rules}
\label{apx:redexreductions}
\begin{figure*}
	\begin{center}
		$
		\begin{array}{rcl}
			\angles{v, \sigma} & \leadsto & \angles{\sigma(v), \sigma} \\ 
			\angles{\text{deref } p, \sigma} & \leadsto & \angles{\sigma(p), \sigma} \\ 
			\angles{c_0[c_1], \sigma} & \leadsto & \angles{\sem{c_0[c_1] : ref}\sigma, \sigma} \\
			\angles{\texttt{new }t[c], \sigma} & \leadsto & \angles{p, \sigma[p \mapsto \texttt{new }t[c] : \text{ref}]} \\
			\angles{\texttt{new hashtable}[t_0,t_1], \sigma} & \leadsto & \angles{p, \sigma[p \mapsto \texttt{new hashtable}[t_0,t_1] : \text{ref}]} \\
			\angles{\text{uop } c, \sigma} & \leadsto & \angles{\sem{\text{uop }c}\sigma, \sigma} \\
			\angles{c_0 \text{ bop } c_1, \sigma} & \leadsto & \angles{\sem{c_0 \text{ bop } c_1}\sigma, \sigma} \\
			\angles{\texttt{true} ? e_0 : e_1, \sigma} & \leadsto & \angles{e_0, \sigma} \\
			\angles{\texttt{false} ? e_0 : e_1, \sigma} & \leadsto & \angles{e_1, \sigma} \\
			\angles{c.\text{fieldname}, \sigma} & \leadsto & \angles{\sem{c.\text{fieldname}}\sigma, \sigma} \\
			\\
			\angles{lval = c, \sigma} & \leadsto & \angles{\sskip, \sigma[lval \mapsto c]} \\
			\angles{t\text{ }v = c, \sigma} & \leadsto & \angles{\sskip, \sigma[v \mapsto c]} \\
			\angles{\texttt{if(true)}\block{s_0}\texttt{else}\block{s_1}, \sigma} & \leadsto & \angles{s_0, \sigma} \\
			\angles{\texttt{if(false)}\block{s_1}\texttt{else}\block{s_1}, \sigma} & \leadsto & \angles{s_1, \sigma} \\
			\angles{\texttt{for}(v\text{-}decls;e?;s_0?) \block{s_1}, \sigma} & \leadsto & \angles{v\text{-}decls; \texttt{while}(e)\block{s_1; s_0}, \sigma}^\dagger \\
			\angles{\texttt{while}(e) \block{s}, \sigma} & \leadsto & \angles{\texttt{if}(e)\block{s; \texttt{while}(e)\block{s}}\texttt{else}\block{\sskip}, \sigma} \\
			\angles{\sskip; s, \sigma} & \leadsto & \angles{s, \sigma} \\
			\\
			\angles{\texttt{commute(false)}\block{\block{s_0} \block{s_1}}, \sigma} & \leadsto & \angles{s_0; s_1, \sigma} \\
		\end{array}
		$
	\end{center}
	\caption{Redex reductions.\\$\dagger$: To satisfy syntax in $v$-$decls$, replace , with ; and $\epsilon$ with skip.}
	\label{Redex rules}
\end{figure*}

Reductions are of the form $\langle r, \sigma \rangle \leadsto \langle \text{r'}, \sigma' \rangle $, where $r'$ is the reduced expression (not necessarily another redex). See Figure \ref{Redex rules} for a full list of redex reductions.
\section{Omitted reduction rules}
\label{apx:omitted}

\subsection{Sequential Semantics $seq$}

Lift the previously defined semantics.

$$
\infer[\text{Lift-Seq}]
{\angles{s, \sigma} \leadsto_{seq} \angles{s', \sigma'}}
{\angles{s, \sigma} \leadsto \angles{s', \sigma'}}
$$

Add the redex:
\[\begin{array}{lll}
	r_{seq} & ::= & \texttt{commute(true)}\block{\block{s_1} \block{s_2}} 
\end{array}\]

With the redex reduction:
\[\begin{array}{lll}
	\angles{\texttt{commute(true)}\block{\block{s_0} \block{s_1}}, \sigma} & \leadsto_{seq} & \angles{s_0; s_1, \sigma}
\end{array}\]

And corresponding small step rule:
$$
\infer[\text{Small-Step-Seq}]
{\angles{H[r_{seq}], \sigma} \leadsto_{seq} \angles{H[r_{seq}'], \sigma'}}
{\angles{r_{seq}, \sigma} \leadsto_{seq} \angles{r_{seq}', \sigma'}}
$$

\subsection{Nondeterministic Semantics $nd$}

Like $seq$, for $nd$, we similarly extend redexes:
\[\begin{array}{lll}
	r_{nd} & ::= & \texttt{commute(true)}\block{\block{s_1},\block{s_2}} 
\end{array}\]
and lift the redex reductions into the $\leadsto_{nd}$ reductions.
Much like the sequential case, lift the redex semantics:
\[\infer[\text{Lift-Nd}]
{\angles{s, \sigma} \leadsto_{nd} \angles{s', \sigma'}}
{\angles{s, \sigma} \leadsto \angles{s', \sigma'}}
\]
And corresponding small step rule:
$$
\infer[\text{Small-Step-Nd}]
{\angles{H[r_{nd}], \sigma} \leadsto_{nd} \angles{H[r_{nd}'], \sigma'}}
{\angles{r_{nd}, \sigma} \leadsto_{nd} \angles{r_{nd}', \sigma'}}
$$

\subsection{Parallel Semantics $par$}

Lifting the redex semantics:
$$
\infer[\text{Lift-Par}]
{\angles{s, \sigma} \leadsto_{par} \angles{s', \sigma'}}
{\angles{s, \sigma} \leadsto \angles{s', \sigma'}}
$$

\subsection{Properties of the Given Semantics}
\label{apx:properties}

We assert that the separation of a statement into a redex and a context is unique:
\begin{lemma}[Redex Uniqueness]
	\label{Uniqueness Proposition}
	$\forall s: \exists! H, r: s = H[r]$
	\begin{proof}
		Proof via analysis of the grammatical structure of $H$. Note that the nonterminal $H$ always occurs on the left, or to the right of a constant, which is not reducible. Likewise, every redex is syntactically distinct.
	\end{proof}
\end{lemma}

Note that this does not entail determinism, as there are multiple ways to reduce some redexes and concurrent configurations. However, it does mean the syntax upon which the next step is performed is well-defined. In the absence of commute structures, we are able to assert determinism:

\begin{lemma}[Conditional Determinism]
	\label{Conditional Determinism Proposition}
	$\forall s, \sigma, H, r: s = H[r] \wedge
	(\forall s_0, s_1: r \neq \texttt{commute(true)} \block{\block{s_0} \block{s_1}}) \rightarrow
	\exists s', \sigma', s'', \sigma'': (\angles{s, \sigma} \leadsto_{sem} \angles{s', \sigma'}) \wedge (\angles{s, \sigma} \leadsto_{sem} \angles{s'', \sigma''}) \rightarrow (s' = s'' \wedge \sigma' = \sigma'')$ 
	\begin{proof}
		Proof via inversion of the redex rules. Each non-\texttt{commute(true)} syntax has a unique rule that can apply to it.
	\end{proof}
\end{lemma}

\begin{lemma}
	\label{Commutativity Correctness}
	Every \texttt{commute} block guard in $s$ is a sufficient commutativity condition $\implies$ $s$ is deterministic in $nd$.
	\begin{proof}
		Recall Definition \ref{Sufficient Commutativity Condition} (\nameref{Sufficient Commutativity Condition}). Proof by induction on \ttt{commute} blocks. If the inside of a \ttt{commute} block is deterministic, and the order does not matter because they commute, then the end state is determined. 
	\end{proof}
\end{lemma}


\begin{lemma}
	\label{Inclusion}
	$\forall s, \sigma: \sem{s}_{seq}(\sigma) \subseteq \sem{s}_{nd}(\sigma) \subseteq \sem{s}_{par}(\sigma)$
	\begin{proof}
		Proof by the construction of executions. We may choose to apply the $\angles{\texttt{commute(true)}\block{\block{s_0} \block{s_1}}, \sigma} \leadsto_{nd} \angles{s_0; s_1, \sigma}$ rule every time, which is identical to the rule in $seq$. Similarly, we may choose to apply Proj-Left (or Proj-Right) to completion to simulate a $com$ trace in $par$.
	\end{proof}
\end{lemma}

\begin{lemma}
	\label{Deterministic Sequential}
	$s$ is deterministic in $sem$ $\implies$ $\sem{s}_{sem} = \sem{s}_{seq}$
	\begin{proof}
		Proof: Note that $s$ is always deterministic in $seq$, as redex reductions are all unique. By Lemma \ref{Inclusion}, $\forall \sigma: \sem{s}_{sem} \supseteq \sem{s}_{seq}$. But they are both singleton sets. Thus they must be equal.
	\end{proof}
\end{lemma}

\subsection{Serializability adapted to nested \ttcommute{} blocks}
\label{apx:adaptedser}
Traditional definitions of serializability adapted to our setting may be defined as something like so:
\begin{definition}[Serial Execution]
	An execution $\execution$ is serial when:
\[
\begin{array}{ll}
	\multicolumn{2}{l}{\forall p, p' \in \{L_n, R_n \mid n \in \mathbb{N}\}^*: } \\ 
	\multicolumn{2}{l}{p \text{ is not a prefix or extension of } p' \implies} \\
	& (\forall \transition, \transition' \in \execution: \transition.\mathit{fr} = p \wedge \\ 
	& \transition'.\mathit{fr} = p' \implies \transition \le_\execution \transition') \vee \\ 
	& (\forall \transition, \transition' \in \execution: \transition.\mathit{fr} = p \wedge \\
	& \transition'.\mathit{fr} = p' \implies \transition' \le_\execution \transition)
\end{array}
\]
\end{definition}
To briefly argue for this definition, the assumption is that we are not talking about a thread's children (with which it will necessarily have interleaving). The body asserts that each thread's steps are completely separated from all (non-descendant) threads. As we did with s-serializability, take a execution to be serializable if there is an equivalent execution that is serial, and a program $s$ to be serializable when all its executions are serializable.

\begin{prop}
	$\forall s: RSerializable(s) \implies Serializable(s)$
\end{prop}
This should be clear as for any prefix $p$, we have $pL_{k_0}$ is neither a prefix nor an extension of $pR_{k_1}$. Thus any execution that is s-serializable is serializable, so the same holds for any program.

\section{Locking}
\label{apx:locking}
Augment states to also hold a map $L : \mathbb{N} \rightarrow \mathbb{B}$ representing whether a lock associated with a number is held.
That is, $ \Sigma := (Varnames \sqcup MemLocs \rightharpoonup V) \times L $. Let us suppose that we can distinguish memory locations from bare natural numbers, so that we may use $\sigma(n)$ to unambiguously refer to the acquired state of lock n.

Suppose that if we have the sum of states $\sigma_0 \oplus \sigma_1$, we always try to acquire from the set of locks associated with the right-most state ($\sigma_1$ here). That is, there is only one global set of locks.

Add redexes:
\[\begin{array}{lll}
	r & ::= & \dots \mid lock(n) \mid unlock(n) 
\end{array}\]

Add contexts:
\[\begin{array}{lll}
	H & ::= & \dots \mid lock(H) \mid unlock(H) 
\end{array}\]

Add special rules:
\[
\infer[Lock]
{\angles{lock(n), \sigma} \leadsto_{par} \angles{\sskip, \sigma[n \mapsto \texttt{true}]}}
{\sigma(n) = \texttt{false}}
\]

\[
\infer[Unlock]
{\angles{unlock(n), \sigma} \leadsto_{par} \angles{\sskip, \sigma[n \mapsto \texttt{false}]}}
{}
\]

\section{Proof of the Main Theorem}
\label{apx:proof}

\mainlemma*

\begin{proof}
Recall from Lemma \ref{Inclusion} that $\forall \sigma: \sem{s}_{nd}(\sigma) \subseteq \sem{s}_{par}(\sigma)$. Thus it only remains to show $RSearlizable(s) \implies \forall \sigma: \sem{s}_{nd}(\sigma) \supseteq \sem{s}_{par}(\sigma)$.

Apply induction on the maximum nesting of \texttt{commute} blocks in s (alternatively, the maximum fragment length for transitions in executions on $s$ in $par$). It is clear that if the maximum length of a fragment is $0$, every $par$ execution of never forks, and thus contains no \texttt{commute(true)}. Stronger, by (repeated application of) Lemma \ref{Conditional Determinism Proposition}, there is a unique execution. Furthermore, as the semantics agree in the absence of \texttt{commute(true)}, this is an execution in $nd$. Thus the base case is satisfied.

Suppose the statement holds for all $s$ with \ttt{commute blocks} nested at most $k$ times. Suppose an r-serializable $s$ is given, with \ttt{commute} blocks nested $k+1$ times. Suppose an arbitrary $\sigma$ is given. Consider an execution $\execution$, with $\angles{s, \sigma} \Downarrow_{par} \execution$. Since $s$ is r-serializable, there exists an r-serial execution $\execution'$ with $\angles{s, \sigma} \Downarrow_{par} \execution'$ and $\execution_f.st = \execution'_f.st$. Construct an execution $\execution_{nd}$.

While a transition in $\execution'$ is not that of a \ttt{commute(true)} block, make the same transition in $\execution_{nd}$. 
When $\execution'$ performs a fork, the configuration will become $\angles{(\angles{s_L, \emptyset}, \angles{s_R, \emptyset}), s', \sigma'}$. 
If the next step is a Join, the case is trivial (the block is a no-op). 
Otherwise, the next transition will use either Proj-Left or Proj-Right. 
Suppose it is a Proj-Left. Then by r-serializability, the $s_L$ will execute to completion before and transitions happen on $s_R$. 
In other words, $\angles{(\angles{s_L, \emptyset}, \angles{s_R, \emptyset}), s', \sigma'} \leadsto_{par}^* \angles{(\angles{\sskip, \sigma_L}, \angles{s_R, \emptyset}), s', \sigma'_L}$. 
By applying the rule used in the premise of Proj-Left, obtain an execution $\angles{s_L, \emptyset \oplus \sigma'} \Downarrow_{par} \execution_L$, with $\execution_{L, f}.st = \sigma_L \oplus \sigma'_L$.

Note that $s_L$ is r-serializable; from any trace take only the transition relating to the current fragment, and remove the leading $L_k$; as r-serializabilty held for all prefixes, it is preserved for this set of prefixes that all started with $L_k$. 
Furthermore, $s_L$ contains commute blocks with nesting at most $k$, so by applying the inductive hypothesis, $\sigma_L \oplus \sigma'_L \in \sem{s_L}_{nd}[\emptyset \oplus \sigma']$. 
Thus there exists an execution $\angles{s_L, \emptyset \oplus \sigma'} \Downarrow_{nd} \execution_{nd, L}$. 
To $\execution_{nd}$, apply the rule $$\angles{\texttt{commute(true)}\block{\block{s_L},\block{s_R}}, \sigma} \leadsto_{nd} \angles{s_L; s_R, \sigma}$$ (with context H = $\bullet$ ; $s'$). 
Every redex reduction in $\execution_{nd, L}$ may be applied to $\execution_{nd}$; apply the grammar rule $H = H; s'$ to each context in $\execution_{nd, L}$.
Afterwards, apply the $\sskip; s' \leadsto s'$ transition (which does not change the state).
The left summand of the state may be discarded, as all values declared within the commute block go out of scope.
Reason similarly about the right fragment to append an $\execution_{nd, R}$ to $\execution_{nd}$.
Finally, in $\execution'$, Join is used; as this only discards the local states, which are already lost to scope in $\execution_{nd}$, it may be disregarded.

If Proj-Right was applied first, the case is symmetrical, and merely apply the transition $$\angles{\texttt{commute(true)}\block{\block{s_L} \block{s_R}}, \sigma} \leadsto_{nd} \angles{s_R; s_L, \sigma}$$ instead.
Continue, repeating this process, until the end of $\execution'$.
Note that at every point outside of the \ttt{commute(true)} blocks, the state in $\execution'$ is equal to the state in our constructed $\execution_{nd}$.
Thus $\execution'_f.st = \execution_{nd, f}.st$, so $\sem{s}_{nd}(\sigma) \supseteq \sem{s}_{par}(\sigma)$ and the proof is complete.
\end{proof}

\maintheorem*
\begin{proof}
By Lemma \ref{RSer and ND}, $\sem{s}_{nd} = \sem{s}_{par}$. By Lemma \ref{Commutativity Correctness}, $\sem{s}_{seq} = \sem{s}_{nd}$. Thus $\sem{s}_{seq} = \sem{s}_{par}$.
\end{proof}
\section{Sugaring and Differences from Formal Semantics}
\label{apx:sugaring}

For theoretical or practical reasons, the implemented language in Veracity differs from the formal language given in a few key ways.
\begin{itemize}
	\item{\bf No explicit dereferencing}. We do not allow the programmer ways for explicitly manipulating pointers. Dereferences are performed automatically when evaluating indexing operations.
	\item{\bf Indexing operations as lvals}. We allow the programmer to use nested indexing operations on the left hand side of assign expressions. This can be seen as syntactic sugar for indexing into a temporary reference, then assigning to that reference.
	\item{\bf IO}. For practical purposes, it is useful to allow for IO operations. This includes both reading/writing to \ttt{stdout}/\ttt{stdin}, and also file operations. 
	\item{\bf Function calls}. Veracity implements a call stack and allows for Veracity subroutines to be called. Semantically, this is not dissimilar to inlining the body in a new scope. None of our benchmarks used recursion, and the behavior of \ttt{commute} blocks in recursive functions is undefined.
	\item{\bf Commute seq/par}. For testing purposes, Veracity \ttt{commute} statements are written as \ttt{commute\_seq} or \ttt{commute\_par}. The former corresponds to the semantics in $nd$, and the latter to the semantics in $par$.
\end{itemize}
\vfill\pagebreak
\section{Loop Summary}
\label{apx:loopsum}

We document how we summarized loops using a combination of Korn and manual reasoning.  We list loop excerpts from benchmarks with loops, with their summary in conjunctive normal form.

\begin{enumerate}
 \item {\bf Benchmark: {\texttt loop-amt.vcy}}
 \begin{enumerate}
     \item {\bf Loop:} \\
     \begin{program}[style=tt]
        while \tab(i > 0) \{
        if (i \% 2 == 0) \{ amt = amt * -1; \}
        else \{ amt = amt * i * -1; \}
        i = i - 1; \}
     \end{program}
     \\ {\bf Summary:} \\
        \begin{program}[style=tt]
            i == 0 $\wedge$ (amt == $amt_{old}$ $\vee$ $i_{old}$ \% 2 $\neq$ 0)
        \end{program}
        
 \end{enumerate}
 \item {\bf Benchmark: {\texttt loop-disjoint.vcy}}
 \begin{enumerate}
    \item {\bf Loop:} \\
    \begin{program}[style=tt]
        assume (x > 0);
        while (x > 0) \{ x = x - 1; \}
    \end{program}
    \\ {\bf Summary:} \\
       \begin{program}[style=tt]
        x == 0
       \end{program}
    \item {\bf Loop:} \\
    \begin{program}[style=tt]
        assume (y > 0);
        while (y > 0) \{ y = y - 1; \}
    \end{program}
    \\ {\bf Summary:} \\
    \begin{program}[style=tt]
        y == 0
    \end{program}
\end{enumerate}
 \item {\bf Benchmark: {\texttt loop-inter.vcy}}
 \begin{enumerate}
    \item {\bf Loop:} \\
    \begin{program}[style=tt]
        assume (x > 0);
        while (x > 0) \{ x = x - 1; s1 = s1 + 1; \}
    \end{program}
    \\ {\bf Summary:} \\
       \begin{program}[style=tt]
            x == 0 $\wedge$ s1 == $s1_{old}$ + $x_{old}$
       \end{program}
    \item {\bf Loop:} \\
    \begin{program}[style=tt]
        assume (y > 0);
        while (y > 0) \{ y = y - 1; s2 = s2 + x; \}
    \end{program}
    \\ {\bf Summary:} \\
        \begin{program}[style=tt]
            y == 0 $\wedge$ $x_{old}$ == x $\wedge$ s2 == $s2_{old}$ + ($x_{old}$ * $y_{old}$)\\
        \end{program}
\end{enumerate}
 \item {\bf Benchmark: {\texttt loop-simple.vcy}}
 \begin{enumerate}
    \item {\bf Loop:} \\
    \begin{program}[style=tt]
        while (y > 0) \{ x = x + 1; y = y - 1; \}
    \end{program}
    \\ {\bf Summary:} \\
       \begin{program}[style=tt]
            y == 0 $\wedge$ x == $x_{old}$ + $y_{old}$
       \end{program}
    \item {\bf Loop:} \\
    \begin{program}[style=tt]
        while (z > 0) \{ x = x + 1; z = z - 1; \}
    \end{program}
    \\ {\bf Summary:} \\
        \begin{program}[style=tt]
            z == 0 $\wedge$ x == $x_{old}$ + $z_{old}$
        \end{program}
\end{enumerate}
\end{enumerate}

\vfill
\section{Unabridged inference benchmarks}
\label{apx:unabridgedinferences}

{\renewcommand{\arraystretch}{\benchtablerowstretch}\footnotesize
\begin{tabularx}{\columnwidth}{|l|r|X|}
	\hline
	Program & Time (s) & Inferred Conditions \\
	\hline
	\hline
	\texttt{array-disjoint} & 0.63 & \texttt{i != j \&\& x != y \textbar\textbar\  x == y}\\
	\texttt{array1} & 0.75 & \texttt{1 != r[0] \&\& r[0] + 1 != y \&\& r[0] <= 1 \textbar\textbar\  r[0] + 1 == y \&\& r[0] <= 1}\\
	\texttt{array2} & 1.11 & \texttt{0 \textgreater\ a[0] \&\& 1 != x \textbar\textbar\  1 == x}\\
	\texttt{array3} & 1.13 & \texttt{d != e \&\& a != b \textbar\textbar\  a == b}\\
	\texttt{calc} & 120.10 & \texttt{1 == y \&\& 0 != y \&\& 1 \textgreater\ c \&\& 1 != c \textbar\textbar\  0 == y \&\& 1 \textgreater\ c \&\& 1 != c \textbar\textbar\  1 == c}\\
	\texttt{conditional} & 0.18 & \texttt{x \textgreater\ 0}\\
	\texttt{counter} & 0.20 & \texttt{0 != c}\\
	\texttt{dict} & 3.82 & \texttt{i != r \&\& c + x != y \textbar\textbar\  c + x == y}\\
	\texttt{dot-product} & 0.24 & \texttt{true}\\
	\texttt{even-odd} & 1.18 & \texttt{x \% 2 == x + y \&\& 0 != y \textbar\textbar\  0 == y}\\
	\texttt{ht-add-put} & 2.24 & \texttt{tbl[z] == u + 1 \&\& u + 1 != z}\\
	\texttt{ht-cond-mem-get} & 1.54 & \texttt{tbl[x] == tbl[z] \&\& x != z \textbar\textbar\  x == z}\\
	\texttt{ht-cond-size-get} & 0.83 & \texttt{ht\_size(tbl) <= 0 \&\& 0 != z \textbar\textbar\  0 == z}\\
	\texttt{ht-simple} & 30.64 & \texttt{x + a != z \&\& 3 == tbl[z] \&\& y != z}\\
	\texttt{linear-bool} & 3.62 & \texttt{0 <= y \&\& 3 == x \&\& 2 != x \&\& 1 != x \&\& x \textgreater\ 0 \&\& 0 != x \textbar\textbar\  0 \textgreater\ y + 3 * x \&\& 2 == x \&\& 1 != x \&\& x \textgreater\ 0 \&\& 0 != x}\\
	\texttt{linear-cond} & 2.65 & \texttt{2 <= y \&\& 2 != y \&\& 1 != y \textbar\textbar\  0 == y \&\& x + 2 == z \&\& 2 \textgreater\ y \&\& 2 != y \&\& 1 != y \textbar\textbar\  2 == y \&\& 1 != y \textbar\textbar\  1 == y}\\
	\texttt{linear} & 0.25 & \texttt{true}\\
	\texttt{loop-amt} & 120.06 & \texttt{0 == i \&\& amt == i\_pre \&\& ctr - 1 \textgreater\ i\_pre \&\& i\_pre <= amt \&\& 0 != i\_pre \&\& i\_pre <= ctr \&\& amt != amt\_pre \&\& ctr - 1 \textgreater\ amt\_pre \&\& amt\_pre <= amt \&\& 0 != amt\_pre \&\& amt\_pre <= ctr \&\& ctr - 1 != 1 \&\& 1 != ctr \&\& 1 != amt \&\& 1 == ctr + amt \textbar\textbar\  0 == i \&\& i\_pre \textgreater\ amt \&\& 0 != i\_pre \&\& i\_pre <= ctr \&\& amt != amt\_pre \&\& ctr - 1 \textgreater\ amt\_pre \&\& amt\_pre <= amt \&\& 0 != amt\_pre \&\& amt\_pre <= ctr \&\& ctr - 1 != 1 \&\& 1 != ctr \&\& 1 != amt \&\& 1 == ctr + amt \textbar\textbar\  i == i\_pre \&\& 0 == i\_pre \&\& i\_pre <= ctr \&\& amt != amt\_pre \&\& ctr - 1 \textgreater\ amt\_pre \&\& amt\_pre <= amt \&\& 0 != amt\_pre \&\& amt\_pre <= ctr \&\& ctr - 1 != 1 \&\& 1 != ctr \&\& 1 != amt \&\& 1 == ctr + amt \textbar\textbar\  0 == i \&\& i <= ctr \&\& i\_pre \textgreater\ amt \&\& amt != i\_pre \&\& 1 != i\_pre \&\& i\_pre \textgreater\ 0 \&\& 0 != i\_pre \&\& i\_pre \textgreater\ ctr \&\& amt != amt\_pre \&\& ctr - 1 \textgreater\ amt\_pre \&\& amt\_pre <= amt \&\& 0 != amt\_pre \&\& amt\_pre <= ctr \&\& ctr - 1 != 1 \&\& 1 != ctr \&\& 1 != amt \&\& 1 == ctr + amt \textbar\textbar\  0 == i \&\& amt == amt\_pre \&\& ctr - 1 \textgreater\ amt\_pre \&\& amt\_pre <= amt \&\& 0 != amt\_pre \&\& amt\_pre <= ctr \&\& ctr - 1 != 1 \&\& 1 != ctr \&\& 1 != amt \&\& 1 == ctr + amt \textbar\textbar\  0 == i \&\& amt\_pre \textgreater\ amt \&\& 0 != amt\_pre \&\& amt\_pre <= ctr \&\& ctr - 1 != 1 \&\& 1 != ctr \&\& 1 != amt \&\& 1 == ctr + amt \textbar\textbar\  amt\_pre == i \&\& 0 == amt\_pre \&\& amt\_pre <= ctr \&\& ctr - 1 != 1 \&\& 1 != ctr \&\& 1 != amt \&\& 1 == ctr + amt \textbar\textbar\  0 == i \&\& i <= ctr \&\& amt != i\_pre \&\& ctr - 1 \textgreater\ i\_pre \&\& i\_pre <= amt \&\& 0 != i\_pre \&\& i\_pre <= ctr \&\& amt\_pre \textgreater\ amt \&\& amt != amt\_pre \&\& 1 != amt\_pre \&\& amt\_pre \textgreater\ 0 \&\& 0 != amt\_pre \&\& amt\_pre \textgreater\ ctr \&\& ctr - 1 != 1 \&\& 1 != ctr \&\& 1 != amt \&\& 1 == ctr + amt \textbar\textbar\  0 == i \&\& amt == i\_pre \&\& ctr - 1 \textgreater\ i\_pre \&\& i\_pre <= amt \&\& 0 != i\_pre \&\& i\_pre <= ctr \&\& amt\_pre \textgreater\ amt \&\& amt != amt\_pre \&\& 1 != amt\_pre \&\& amt\_pre \textgreater\ 0 \&\& 0 != amt\_pre \&\& amt\_pre \textgreater\ ctr \&\& ctr - 1 != 1 \&\& 1 != ctr \&\& 1 != amt \&\& 1 == ctr + amt \textbar\textbar\  0 == i \&\& i\_pre \textgreater\ amt \&\& 0 != i\_pre \&\& i\_pre <= ctr \&\& amt\_pre \textgreater\ amt \&\& amt != amt\_pre \&\& 1 != amt\_pre \&\& amt\_pre \textgreater\ 0 \&\& 0 != amt\_pre \&\& amt\_pre \textgreater\ ctr \&\& ctr - 1 != 1 \&\& 1 != ctr \&\& 1 != amt \&\& 1 == ctr + amt \textbar\textbar\  i == i\_pre \&\& 0 == i\_pre \&\& i\_pre <= ctr \&\& amt\_pre \textgreater\ amt \&\& amt != amt\_pre \&\& 1 != amt\_pre \&\& amt\_pre \textgreater\ 0 \&\& 0 != amt\_pre \&\& amt\_pre \textgreater\ ctr \&\& ctr - 1 != 1 \&\& 1 != ctr \&\& 1 != amt \&\& 1 == ctr + amt \textbar\textbar\  0 == i \&\& i <= ctr \&\& i\_pre \textgreater\ amt \&\& amt != i\_pre \&\& 1 != i\_pre \&\& i\_pre \textgreater\ 0 \&\& 0 != i\_pre \&\& i\_pre \textgreater\ ctr \&\& amt\_pre \textgreater\ amt \&\& amt != amt\_pre \&\& 1 != amt\_pre \&\& amt\_pre \textgreater\ 0 \&\& 0 != amt\_pre \&\& amt\_pre \textgreater\ ctr \&\& ctr - 1 != 1 \&\& 1 != ctr \&\& 1 != amt \&\& 1 == ctr + amt \textbar\textbar\  0 == i \&\& ctr - 1 == 1 \&\& 1 != ctr \&\& 1 != amt \&\& 1 == ctr + amt \textbar\textbar\  amt == i \&\& 1 == ctr \&\& 1 != amt \&\& 1 == ctr + amt}\\
\end{tabularx}}

{\renewcommand{\arraystretch}{\benchtablerowstretch}\footnotesize
\begin{tabularx}{\columnwidth}{|l|r|X|}
	\hline
   (continued) && \\
	\texttt{loop-disjoint} & 0.02 & \texttt{true}\\
	\texttt{loop-inter} & 4.63 & \texttt{0 == x \&\& 0 != y \textbar\textbar\  0 == y}\\
	\texttt{loop-simple} & 0.06 & \texttt{true}\\
	\texttt{matrix} & 0.71 & \texttt{0 == y}\\
	\texttt{nested-counter} & 6.25 & \texttt{0 != c \&\& c != t \textbar\textbar\  c == t; c != x \&\& c <= x \&\& 1 != x \&\& t == x \textbar\textbar\  c \textgreater\ 1 \&\& 0 != c \&\& c == x \&\& c <= x \&\& 1 != x \&\& t == x \textbar\textbar\  1 == x \&\& t == x}\\
	\texttt{nested} & 0.25 & \texttt{true; 0 == x}\\
	\texttt{nonlinear} & 1.42 & \texttt{0 == y}\\
	\texttt{simple} & 3.49 & \texttt{a <= c \&\& a != b \&\& a != c \textbar\textbar\  a == b \&\& a != c}\\
	\hline
\end{tabularx}}
\vfill\pagebreak
\onecolumn
\lstset{tabsize=2}
\section{Source code of benchmark programs}

\label{apx:benchmarks}

{\bf Benchmark: \texttt{array-disjoint.vcy}}
\begin{quote}
\begin{lstlisting}[language=VCY,style=vcy]
int main(int argc, string[] argv) {
	int[] a = new int[5];

	int c = int_of_string(argv[1]);

	int i = int_of_string(argv[2]);
	int j = int_of_string(argv[3]);

	int x = int_of_string(argv[4]);
	int y = int_of_string(argv[5]);

	commute _ {
		{ a[i] = x; busy_wait(c);}
		{ a[j] = y; busy_wait(c);}
	}

	return 0;
}
\end{lstlisting}
\end{quote}

{\bf Benchmark: \texttt{array1.vcy}}
\begin{quote}
\begin{lstlisting}[language=VCY,style=vcy]
int main(int argc, string[] argv) {
    int n = int_of_string(argv[1]);
    int x = int_of_string(argv[2]);
    int y = int_of_string(argv[3]); 
    int[] r = new int[1];
    r[0] = x;
    
    
    commute _ {
        { r[0] = r[0] + 1; busy_wait(n); } 
        { if(r[0] > 1) { r[0] = y; } 
          busy_wait(n);
        }
    }
    
    return 0;
}
\end{lstlisting}
\end{quote}

{\bf Benchmark: \texttt{array2.vcy}}
\begin{quote}
\begin{lstlisting}[language=VCY,style=vcy]

int main(int argc, string[] argv) {
    int n = int_of_string(argv[1]);
    int x = int_of_string(argv[2]);
    int[] a = new int[1];
    a[0] = 69;
    
    commute _ {
        { busy_wait(n); a[0] = a[0] + 1; }
        {  if(a[0] > 0) {a[0] = a[0] * x; } busy_wait(n); }
    }
    
    return 0;
}
\end{lstlisting}
\end{quote}

{\bf Benchmark: \texttt{array3.vcy}}
\begin{quote}
\begin{lstlisting}[language=VCY,style=vcy]
int main(int argc, string[] argv) {
    int a = 0;
    int b = 1;
    int c = 42;
    int d = 0;
    int n = 10;
    int size = int_of_string(argv[1]);
    int[] arr = new int[10];
    
    int e = c % n;
    
    commute _ {
        { busy_wait(size); a = a + 1; arr[e] = b; a = a - 1; }
        { arr[d] = a; busy_wait(size); }
    }
    
    return 0;
}
\end{lstlisting}
\end{quote}

{\bf Benchmark: \texttt{calc.vcy}}
\begin{quote}
\begin{lstlisting}[language=VCY,style=vcy]
int main(int argc, string[] argv) {
    int size = int_of_string(argv[1]);
    int x = 0;
    int y = int_of_string(argv[2]);
    int z = 0;
    int a = int_of_string(argv[3]);
    int c = int_of_string(argv[4]);
    
    commute _ {
        { 
            busy_wait(size);
            x = /* calc1 */ (a);
            c = c + (x*x); 
        } 
        { 
            if (c>0 && y<0)
              { c = c - 1; busy_wait(size); z = /* calc2 */ (y); }
            else
              { busy_wait(size); z = /* calc3 */ (y); }
        }
    }

    return x;
}
\end{lstlisting}
\end{quote}

{\bf Benchmark: \texttt{conditional.vcy}}
\begin{quote}
\begin{lstlisting}[language=VCY,style=vcy]
int main(int argc, string[] argv) {
  int n = int_of_string(argv[1]);
  int x = 42;
  bool r = false;
  commute _ {
    { x = 1; busy_wait(n); }
    { int t = x; busy_wait(n); r = t > 0; }
  }	
  return 0;
}
\end{lstlisting}
\end{quote}

{\bf Benchmark: \texttt{counter.vcy}}
\begin{quote}
\begin{lstlisting}[language=VCY,style=vcy]
int main(int argc, string[] argv) {
	int x = int_of_string(argv[1]);
	int c = 42;

	
	commute _ {
		{ c = c + 1; busy_wait(x);}
		{ if (c > 0) { c = c - 1; busy_wait(x); } else { }}
	}

	print(string_of_int(c) ^ "\n");
	return 0;
}
\end{lstlisting}
\end{quote}

{\bf Benchmark: \texttt{crowdfund.vcy}}
\begin{quote}
\begin{lstlisting}[language=VCY,style=vcy]

void busy(int fuel) {
    while (fuel > 0) {
        fuel = fuel - 1;
    }
    return;
}

void lock(int l) { mutex_lock(l); return; }
void unlock(int l)  { mutex_unlock(l); return; }

void addOrInit(hashtable[int, int] table, int key, int amount) {
    table[key] = lookupDefault(table, key, 0) + amount;
    return;
}

int lookupDefault(hashtable[int, int] table, int key, int v_0) {
    if(ht_mem(table, key)) { return table[key]; }
    else { return v_0; }
}

int f_start_date = 5;
int f_end_date = 10;
int f_goal = 10000;
int f_amount = 15;
int addr_f_escrow = 101;
int addr_f_creator = 102;
int addr_f_receiver = 103;
bool f_CloseRemainderTo = false;
int f_close_date = 9;

int addr_noone = -1;

int x = 0;

/* How much an actor has given */
hashtable[int,int] given = new hashtable[int,int];
hashtable[int,int] balances = new hashtable[int,int];

int donate(int donator, int amount, int nowTS) {
    int r = 0;
    /* Ensures the donation is within the beginning and ending dates of the fund. */
    if (nowTS <= f_start_date || nowTS >= f_end_date) { r = 0; } else {
        /* Verifies that this is a grouped transaction with the second one being a payment to the escrow. */
        lock(donator);
        int curBal = lookupDefault(balances, donator, 0);
        unlock(donator);
        if(curBal < amount) { r = 0; } else {
            lock(donator);
            lock(0);
            lock(addr_f_escrow);
            f_amount = f_amount + amount;
            addOrInit(balances, addr_f_escrow, amount);
            addOrInit(balances, donator, -amount);
            unlock(0);
            unlock(addr_f_escrow);
            addOrInit(given, donator, amount);
            addOrInit(balances, donator, amount);
            unlock(donator);
            busy(x);
            r = 1;
    }   }
    return r;
}

int reclaim(int donator, int amount, int nowTS) {
   if (f_close_date > nowTS) { return 0; }
   if (f_goal < f_amount) { return 0; }

   int fee = 0;
   int t = amount + fee;
   lock(donator);
   int g = given[donator];
   unlock(donator);
   if (t > g) { return 0; }
   if (f_CloseRemainderTo) { return 0; }
   if (! ht_mem(balances, addr_f_escrow)) { return 0; }
   lock(donator);
   lock(addr_f_escrow);
   addOrInit(balances, donator, amount);
   addOrInit(balances, addr_f_escrow, -amount);
   unlock(addr_f_escrow);
   unlock(donator);
   busy(x);
   return 1;
}

int main(int argc, string[] argv) {
    x = int_of_string(argv[1]);
    int nowTS = 7;

    
    mutex_init(addr_f_escrow);
    
    
    
    
    
    
    balances[1] = 1000;
    balances[2] = 1000;
    balances[3] = 1000;
    balances[4] = 1000;
    
    commute {
        { donate(1, 10, nowTS); }
        { donate(2, 10, nowTS); }
        { donate(3, 10, nowTS); }
        { donate(4, 10, nowTS); }
    }
    
    commute {
        { reclaim(1, 10, nowTS); }
        { reclaim(2, 10, nowTS); }
        { reclaim(3, 10, nowTS); }
        { reclaim(4, 10, nowTS); }
    }
    return 0;
}
\end{lstlisting}
\end{quote}

{\bf Benchmark: \texttt{dict.vcy}}
\begin{quote}
\begin{lstlisting}[language=VCY,style=vcy]
int main(int argc, string[] argv) {
	int n = int_of_string(argv[1]);
	int t = random(0, 100);
        int c = 69;
	int r = random(0, 100);
	int y = 42 + 69;
	int x = 42;
	int i = r + 1;
	hashtable[int, int] stats = new hashtable[int, int];
	stats[r] = 0;
	stats[i] = 0;

	commute _ {
		{ busy_wait(n); t = c + x; ht_put(stats,r,t); }
		{ ht_put(stats,i,y); y = y + x; busy_wait(n); } 
	}

	return x;
}
\end{lstlisting}
\end{quote}

{\bf Benchmark: \texttt{dihedral.vcy}}
\begin{quote}
\begin{lstlisting}[language=VCY,style=vcy]
/* https://www.whitman.edu/documents/Academics/Mathematics/SeniorProject_CodyClifton.pdf */

int n = 1;

/* r and s are generators of dihedral group Dn 
 * All rigid motions are representable as r^k or r^k*s
 */

int rk(int a, int k) => (a + k) % n;
int r(int a)         => rk(a, 1);

int s(int a) => (n - a) % n;

int[] id() {
    int[] a = new int[n];
    for (int i = 0; i < n; i = i + 1;) {
        a[i] = i;
    }
    return a;
}

/* Apply rigid motions 
 * Rotation and symmetry operations can be
 *   started at the middle of the array and wrapped around,
 *   instead of starting at the beginning.
 * That way, if two threads are operating on the array,
 *   one can be offset from the other,
 *   making it less likely that they will fight over values.
 */

void mrk(int[] x, int k, bool start_in_middle) {
    int m = start_in_middle ? n / 2 : 0;
    for (int i = m; i < n; i = i + 1;) {
        x[i] = rk(x[i], k);
    }
    for (int i = 0; i < m; i = i + 1;) {
        x[i] = rk(x[i], k);
    }
    return;
}

void mr(int[] x) {
    mrk(x, 1);
    return;
}

void ms(int[] x, bool start_in_middle) {
    int m = start_in_middle ? n / 2 : 0;
    for (int i = m; i < n; i = i + 1;) {
        x[i] = s(x[i]);
    }
    for (int i = 0; i < m; i = i + 1;) {
        x[i] = s(x[i]);
    }
    return;
}


/* Inputs are
 *   n > 1
 *   r1 >= 0
 *   s1 in [0, 1]
 *   r2 >= 0
 *   s2 in [0, 1]
 *  
 * First rigid motion is r^r1 * s^s1
 * Second rigid motion is r^r2 * s^s2
 */

int main(int argc, string[] argv) {
    n = int_of_string(argv[1]);

    int[] a = id();

    /* Rigid motion 1 */
    int r1 = int_of_string(argv[2]) % n;
    bool s1 = int_of_string(argv[3]) > 0;

    /* Rigid motion 2 */
    int r2 = int_of_string(argv[4]) % n;
    bool s2 = int_of_string(argv[5]) > 0;


    /* If s1 and not s2 then commute if
            n is even and r2 = 0 or r2 = n/2
            n is odd and r2 = 0
       If not s1 and s2 then commute if
            n is even and r1 = 0 or r1 = n/2
            n is odd and r1 = 0
       If s1 and s2 then commute if
            r1 = r2
       If not s1 and not s2 then always commute
    */
    bool phi =
        (!s1 && !s2) ||
        (s1 && !s2 && (r2 == 0 || (n % 2 == 0 && r2 == n / 2))) ||
        (!s1 && s2 && (r1 == 0 || (n % 2 == 0 && r1 == n / 2))) ||
        (s1 && s2 && r1 == r2);
    
    /*print(phi ? "true\n" : "false\n");*/

    commute (phi) {
        { mrk(a, r1, true);
          if (s1) { ms(a, true); }
        }
        { mrk(a, r2, false);
          if (s2) { ms(a, false); }
        }
    }

    for (int i = 0; i < n; i = i + 1;) {
        print(string_of_int(a[i]) ^ " ");
    }
    print("\n");

    return (phi ? 1 : 0);
}
\end{lstlisting}
\end{quote}

{\bf Benchmark: \texttt{dot-product.vcy}}
\begin{quote}
\begin{lstlisting}[language=VCY,style=vcy]
/* Dot product of 2D vectors */

int dot_product(int[] x, int[] y, int d) {
	commute _ {
		{ d = d + x[0] * y[0]; busy_wait(d);}
		{ busy_wait(d); d = d + x[1] * y[1]; }
	}
	return d;
}

int main(int argc, string[] argv) {

	int d = int_of_string(argv[1]);
	
	int[] x = new int[] {
		int_of_string(argv[2]),
		int_of_string(argv[3])
	};

	int[] y = new int[] {
		int_of_string(argv[4]),
		int_of_string(argv[5])
	};

	return dot_product(x, y, d);
}
\end{lstlisting}
\end{quote}

{\bf Benchmark: \texttt{even-odd.vcy}}
\begin{quote}
\begin{lstlisting}[language=VCY,style=vcy]

int main(int argc, string[] argv) {
    int n = int_of_string(argv[1]);
    int x = int_of_string(argv[2]);
    int y = int_of_string(argv[3]);

    
    commute _ {
        { busy_wait(n); 
          if (x % 2 == 0){
            x = x + y;
          } 
        }
        { if(x % 2 == 1){
            x = x + y;
          } 
          busy_wait(n);
        }
    } 
    
    return x;
}
\end{lstlisting}
\end{quote}

{\bf Benchmark: \texttt{filesystem.vcy}}
\begin{quote}
\begin{lstlisting}[language=VCY,style=vcy]

int[] data = new int[1024];

int main(int argc, string[] argv) {
    int n = int_of_string(argv[1]);

    hashtable[string, string] ht = new hashtable[string, string];
    
    for(int i = 0; i < 1024; i = i + 1;) {
        data[i] = i;
    }
    
    string fname1 = "test1.txt";
    string fname2 = "test2.txt";
    
    commute (fname1 != fname2) {
        { 
            out_channel fout1 = open_write(fname1);
            for(int i=0; i < n; i = i + 1;) { write(fout1, string_of_int(data[i%1024]) ^ "\n"); } 
            close(fout1);
        }
        { 
            out_channel fout2 = open_write(fname2);
            for(int i=n-1; i >= 0; i = i - 1;) { write(fout2, string_of_int(data[i%1024]) ^ "\n"); } 
            close(fout2);
        }
    }
    
    return 0;
}
\end{lstlisting}
\end{quote}

{\bf Benchmark: \texttt{ht-add-put.vcy}}
\begin{quote}
\begin{lstlisting}[language=VCY,style=vcy]
int main(int argc, string[] argv) {
	hashtable[int,int] tbl = new hashtable_seq[int,int];
	int n = int_of_string(argv[1]);
	int y = int_of_string(argv[2]);
	int z = int_of_string(argv[3]);
	int u = int_of_string(argv[4]);
	
	tbl[z] = u + 1;
	
	commute _ {
		{ 	busy_wait(n);
			y = u + 1;
		  	ht_put(tbl, y, u); 
		}
		{ 	y = ht_get(tbl, z); 
		 	busy_wait(n);
		}
	}
	
	return 0;
}
\end{lstlisting}
\end{quote}

{\bf Benchmark: \texttt{ht-cond-mem-get.vcy}}
\begin{quote}
\begin{lstlisting}[language=VCY,style=vcy]
int main(int argc, string[] argv) {
	hashtable[int,int] tbl = new hashtable[int,int];
	int n = int_of_string(argv[1]);
	int x = int_of_string(argv[2]);
	int y = int_of_string(argv[3]);
	int z = int_of_string(argv[4]);

	tbl[x] = 42;
	tbl[z] = 42;
	
	commute _ {
		{ 
			busy_wait(n); 
			if(ht_mem(tbl, x)) { 
		    	y = ht_get(tbl, x);
		  	}  
		}
		{	if(ht_mem(tbl, z)) {
				y = ht_get(tbl, z);
			} 
			busy_wait(n);}
	}

	return 0;
}
\end{lstlisting}
\end{quote}

{\bf Benchmark: \texttt{ht-cond-size-get.vcy}}
\begin{quote}
\begin{lstlisting}[language=VCY,style=vcy]
int main(int argc, string[] argv) {
	hashtable[int,int] tbl = new hashtable_seq[int,int];
	int n = int_of_string(argv[1]);
	int y = int_of_string(argv[2]);
	int z = int_of_string(argv[3]);

	commute _ {
		{ 	busy_wait(n);
			if(ht_size(tbl) > 0) { 
			y = y + z;
		  	}
		}
		{ 	y = ht_get(tbl,z);
			busy_wait(n);
		}
	}
	
	return 0;
}
\end{lstlisting}
\end{quote}

{\bf Benchmark: \texttt{ht-fizz-buzz.vcy}}
\begin{quote}
\begin{lstlisting}[language=VCY,style=vcy]
void busy(int fuel) {
    while (fuel > 0) { fuel = fuel - 1; }
    return;
}

int main(int argc, string[] argv) {
    int n = int_of_string(argv[1]);

    hashtable[int, int] s = new hashtable[int, int];

    commute {
      { for (int i = n - 1; i > 0; i = i - 2;) {
            s[i] = 1;
        }
      } {
        for (int i = 3; i < n; i = i + 3;) {
            s[i] = 1;
        }
      } {
        for (int i = 5; i < n; i = i + 5;) {
            s[i] = 1;
        }
      }
    }

    int total = 0;

    commute {
        {
            for (int i = 0; i < n / 2; i = i + 1;) {
                total = total + ((s[i] == 1) ? 1 : 0);
            }
        } {
            for (int i = n / 2; i < n; i = i + 1;) {
                total = total + ((s[i] == 1) ? 1 : 0);
            }
        }
    }

    return total;
}
\end{lstlisting}
\end{quote}

{\bf Benchmark: \texttt{ht-simple.vcy}}
\begin{quote}
\begin{lstlisting}[language=VCY,style=vcy]
int main(int argc, string[] argv) {
	hashtable[int,int] tbl = new hashtable[int,int];
	int n = int_of_string(argv[1]);
	int x = int_of_string(argv[2]);
	int y = int_of_string(argv[3]);
	int z = int_of_string(argv[4]);
	int a = int_of_string(argv[5]);
	int v = int_of_string(argv[6]);
	int g = int_of_string(argv[7]);
	
	tbl[z] = 3;
	
	commute _ {
		{ busy_wait(n); 
		  v = ht_get(tbl, z);
		  v = v + 3;
		  g = x + a;
		  ht_put(tbl, g, v); }
		{ v = ht_get(tbl, z);
		  v = v * 2;
		  ht_put(tbl, y, v); 
		  busy_wait(n); }
	}

	return 0;
}
\end{lstlisting}
\end{quote}

{\bf Benchmark: \texttt{linear-bool.vcy}}
\begin{quote}
\begin{lstlisting}[language=VCY,style=vcy]
int main(int argc, string[] argv) {
	int size = int_of_string(argv[1]);
	int x = int_of_string(argv[2]);
	int y = int_of_string(argv[2]);
	int b = 0;
	
	
	
	commute _ {
		{ busy_wait(size); y = y + 3*x; }
		{ if (y < 0) { x = 2; } else { x = 3; } busy_wait(size); }
	}

	return x;
}
\end{lstlisting}
\end{quote}

{\bf Benchmark: \texttt{linear-cond.vcy}}
\begin{quote}
\begin{lstlisting}[language=VCY,style=vcy]
int main(int argc, string[] argv) {
	int size = int_of_string(argv[1]);
	
	
	int x = int_of_string(argv[2]);
	int y = int_of_string(argv[3]);
	int z = int_of_string(argv[4]);

	commute _ {
	    { 
	      busy_wait(size);
		  
		  if (y > 0) {
			  x = x + 2;
		  } else {
			  x = z;
		  }
		  
		}
		{ x = x + 1; y = y + 1; busy_wait(size); }
	}

	print(
		string_of_int(x) ^ " " ^ 
		string_of_int(y) ^ " " ^ 
		string_of_int(z) ^ "\n"
	);

	return 0;
}
\end{lstlisting}
\end{quote}

{\bf Benchmark: \texttt{linear.vcy}}
\begin{quote}
\begin{lstlisting}[language=VCY,style=vcy]
int main(int argc, string[] argv) {
	int size = int_of_string(argv[1]);
	int x = random(-100,100);
	int y = random(-100,100);
	int z = random(-100,100);
	int w = random(-100,100);
	int u = random(-100,100);

	
	commute _ {
		{ x = x + 2*w + u; busy_wait(size); }
		{ busy_wait(size); x = x + y + 3*z; }
	}
	
	return 0;
}
\end{lstlisting}
\end{quote}

{\bf Benchmark: \texttt{loop-amt.vcy}}
\begin{quote}
\begin{lstlisting}[language=VCY,style=vcy]
int main(int argc, string[] argv) {
    int i = int_of_string(argv[1]);
    int ctr = int_of_string(argv[2]);
    int amt = 2;
    int n = i;
    
    commute _ {
        {
            /* this only works if i is initially even. */
            while(i>0) { 
                if(i%2 == 0) {
                    amt = amt * -1;
                } else {
                    amt = amt * i * -1;
                }
                i = i - 1;
            }
            /* need a strong enough loop invariarnt to show that amt>0 here */
            ctr = ctr + amt; 
        }
        {   
            if (ctr > 0) { ctr = ctr - 1; } busy_wait(n);
        }
    }
    
    return 0;
}
\end{lstlisting}
\end{quote}

{\bf Benchmark: \texttt{loop-disjoint.vcy}}
\begin{quote}
\begin{lstlisting}[language=VCY,style=vcy]
int main(int argc, string[] argv) {
    int x = int_of_string(argv[1]);
    int y = int_of_string(argv[2]);
    
    havoc x; assume(x>0);
    havoc y; assume(y>0);

    commute _ {
        {
            while(x>0) { x = x - 1; }
        }
        {
            while(y>0) { y = y - 1; }
        }
    }
    return 0;
}
\end{lstlisting}
\end{quote}

{\bf Benchmark: \texttt{loop-inter.vcy}}
\begin{quote}
\begin{lstlisting}[language=VCY,style=vcy]
int main(int argc, string[] argv) {
    int x = int_of_string(argv[1]);
    int y = x;
    int s1 = 0;
    int s2 = 0;
    int acc = 0;
    
    havoc x; assume(x>0);
    havoc y; assume(y>0);
    
    havoc s1;
    havoc s2;
    commute _ {
        {
            while(x>0) { x = x - 1; s1 = s1 + 1; }
            acc = acc + s1;
        }
        {
            while(y>0) { y = y - 1; s2 = s2 + x; }
            acc = acc + s2;
        }
    }
    return 0;
}
\end{lstlisting}
\end{quote}

{\bf Benchmark: \texttt{loop-simple.vcy}}
\begin{quote}
\begin{lstlisting}[language=VCY,style=vcy]
int main(int argc, string[] argv) {
    int y = int_of_string(argv[1]);
    int z = int_of_string(argv[2]);
    int x = 0;
    
    havoc x;
    havoc y;
    havoc z;
    
    commute _ {
        {
            while(y>0) { x = x + 1; y = y - 1; } 
        }
        {
            while(z>0) { x = x + 1; z = z - 1; } 
        }
    }
    return 0;
}
\end{lstlisting}
\end{quote}

{\bf Benchmark: \texttt{matrix.vcy}}
\begin{quote}
\begin{lstlisting}[language=VCY,style=vcy]
int main(int argc, string[] argv) {
    int size = int_of_string(argv[1]);
    int s = 0;
    int x = int_of_string(argv[2]);
    int y = int_of_string(argv[3]);
    int z = int_of_string(argv[4]);
    int yy = 0;
    int out = 0;
    commute _ {
        {
          yy = y; busy_wait(size); s = yy;
          x = s*x; y = 3*yy;
          z = z + 2*yy;
          s = 0; yy = 0;
        } 
        { 
          yy = y;
          x = 5*x; y = 4*yy; 
          z = 3*z - yy;
          busy_wait(size); out = z;
          yy = 0;
        }
    }

    return x;
}
\end{lstlisting}
\end{quote}

{\bf Benchmark: \texttt{nested-counter.vcy}}
\begin{quote}
\begin{lstlisting}[language=VCY,style=vcy]
int main(int argc, string[] argv) {
    int n = int_of_string(argv[1]);
    int x = 500;
    int y = random(0, 100);
    int c = 420;
    int t = 500;
    

    commute _  /* 2nd Condition */ {
        { commute _ /* 1st Condition */ { 
              { c = c + 1; } 
              { /*t=*/ busy_wait(n); if(c>0) { c = c - t; } }
          } }
        { t = x;
          busy_wait(n); 
          if(c>t) { x = x - c; } }
    }

    return x;
}
\end{lstlisting}
\end{quote}

{\bf Benchmark: \texttt{nested.vcy}}
\begin{quote}
\begin{lstlisting}[language=VCY,style=vcy]
int main(int argc, string[] argv) {
	int w = int_of_string(argv[1]);
	int x = 0;
	int y = random(0, 100);
	
	commute _ {
		{ commute _ { 
			  { x = 0; busy_wait(w); }
			  { x = x * 2; }
		  } }
		{ y = x; busy_wait(w); }
	}

	return x;
}
\end{lstlisting}
\end{quote}

{\bf Benchmark: \texttt{nonlinear.vcy}}
\begin{quote}
\begin{lstlisting}[language=VCY,style=vcy]
int main(int argc, string[] argv) {
	int n = int_of_string(argv[1]);
	int x = int_of_string(argv[2]);
	int z = int_of_string(argv[3]);
	int w = int_of_string(argv[4]);
	int y = int_of_string(argv[5]);
	int t = 0;
	int u = 0;
	int o = 0;
	int r = random(0, 100);
	
	/* to parallelize the busy's, need commutativity
	   to know the order can be swapped */
	commute _ {
		{ t = r; /* t = */ busy_wait(n);
		  
		  z = z * t;
		  y = 2 * y;
		  x = x + y; 
		  }
		{ 
		  z = 3 * z;
		  x = x + y; 
		  u = x; 
		  
		  busy_wait(n); o = u /* busy_wait(n, u) */; }
	}
	
	return 0;
}
\end{lstlisting}
\end{quote}

{\bf Benchmark: \texttt{simple.vcy}}
\begin{quote}
\begin{lstlisting}[language=VCY,style=vcy]
void compute1(bool cond){} 
void compute2(bool cond){}

int main(int argc, string[] argv) {
    int n = int_of_string(argv[1]);
    int t = int_of_string(argv[2]);
    int a = int_of_string(argv[3]);
    int b = int_of_string(argv[4]);
    int c = int_of_string(argv[5]);
    int u = int_of_string(argv[6]);
    int temp = 0;
    
    commute _ {
        {
            t = /*compute1(c > b);*/
                ((c > b) ? 2 : 1); busy_wait(n);
            a = a - ((t < 0) ? -t : t); 
        }
        {
            u = /*compute2(c > a);*/
                ((c > a) ? 2 : 1); busy_wait(n);
        }
    }
    return 0;
}
\end{lstlisting}
\end{quote}

\vfill
\pagebreak
\section{Using Snapshots (Sec.~\ref{subsec:enforcing}) to avoid locks}

{\bf Result of using snapshots to avoid the need for locks for 
benchmark: \texttt{array3.vcy}}
\begin{quote}
\begin{lstlisting}[language=VCY,style=vcy]
int main(int argc, string[] argv) {
    int a = 0;
    int b = 1;
    int c = 42;
    int d = 0;
    int n = 10;
    int size = int_of_string(argv[1]);
    int[] arr = new int[10];
    
    int e = c % n;

    if (!(d == e) && !(a == b) || a == b) {
      int a0 = a;
      commute (!(d == e) && !(a == b) || a == b) {
        { busy_wait(size); a = a + 1; arr[e] = b; a = a - 1; }
        { arr[d] = a0; busy_wait(size); }
      }
    } else {
        busy_wait(size); a = a + 1; arr[e] = b; a = a - 1;
        arr[d] = a; busy_wait(size);
    } 
   
    return 0;
}
\end{lstlisting}
\end{quote}

{\bf Result of using snapshots to avoid the need for locks for 
benchmark: \texttt{simple.vcy}}
\begin{quote}
\begin{lstlisting}[language=VCY,style=vcy]
void compute1(bool cond){} 
void compute2(bool cond){}

int main(int argc, string[] argv) {
    int n = int_of_string(argv[1]);
    int t = int_of_string(argv[2]);
    int a = int_of_string(argv[3]);
    int b = int_of_string(argv[4]);
    int c = int_of_string(argv[5]);
    int u = int_of_string(argv[6]);
    if(c>a) {
        int a0 = a;
        commute (c>a) {
            {
                t = /*compute1(c > b);*/
                    ((c > b) ? 2 : 1); busy_wait(n);
                a = a - ((t < 0) ? -t : t);
            }
            {
                u = /*compute2(c > a);*/
                    ((c > a0) ? 2 : 1); busy_wait(n);
            }
        }
    } else {
            t = /*compute1(c > b);*/
                ((c > b) ? 2 : 1); busy_wait(n);
            a = a - ((t < 0) ? -t : t);
            u = /*compute2(c > a);*/
                ((c > a) ? 2 : 1); busy_wait(n);
    }
    return 0;
}
\end{lstlisting}
\end{quote}

\end{document}